\documentclass{jnmp01b}
\usepackage{amsmath}

\numberwithin{equation}{section}
\allowdisplaybreaks
\theoremstyle{definition}
\input{tcilatex}
\begin{document}

%\begin{abstract}
%\noindent
%A review of a new separability theory based on degenerated Poisson pencils
%and the so-called separation curves is presented. This theory can be
%considered as an alternative to the Sklyanin theory based on Lax
%representations and the so-called spectral curves.
%\end{abstract}

\renewcommand{\evenhead}{M.\ B\l aszak} 
\renewcommand{\oddhead}{Degenerate Poisson Pencils on Curves:
New Separability Theory}

% Title

\thispagestyle{empty}

\begin{flushleft}
\footnotesize \sf
Journal of Nonlinear Mathematical Physics \qquad 2000, V.7, N~2,
\pageref{firstpage}--\pageref{lastpage}.
\hfill {\sc Review Article}
\end{flushleft}

\vspace{-5mm}

\copyrightnote{2000}{M.\ B\l aszak}

\Name{Degenerate Poisson Pencils on Curves:\newline
New Separability Theory}

\label{firstpage}

\Author{Maciej B\L ASZAK}

\Adress{Physics Department, A. Mickiewicz University, Umultowska 85, 61-614
Poznan, Poland\newline
e-mail: blaszakm@main.amu.edu.pl}

\Date{Received September 29, 1999; Revised February 1, 2000; Accepted
February 2, 2000}

\begin{abstract}
\noindent
A review of a new separability theory based on degenerated Poisson pencils
and the so-called separation curves is presented. This theory can be
considered as an alternative to the Sklyanin theory based on Lax
representations and the so-called spectral curves.
\end{abstract}

% The article

\section{Introduction}

The separation of variables belongs to the basic methods of mathematical
physics from the previous century. Originating from the early works of
D'Alambert (the $18$\emph{th} century), Fourier and Jacobi (the first half
of the $19$\emph{th} century), for many decades it has been the only known
method of exact solution of dynamical systems. Let us briefly recall the
idea from classical mechanics. Consider a Hamiltonian mechanical system of $%
2n$ degrees of freedom and integrable in the sense of Liouville/Arnold
theorem. This means that there exist $n$ linearly independent functions $h_i$%
(Hamiltonians) which are in involution with respect to the canonical Poisson
bracket 
\begin{equation}
\{h_j,h_k\}=0,\,\,\,\,\,\,\,\,\,j,k=1,...,n.  \label{1}
\end{equation}
A system of canonical variables $(\mu _i,\lambda _i)_{i=1}^n$%
\begin{equation}
\{\lambda _i,\lambda _j\}=\{\mu _i,\mu _j\}=0,\,\,\,\,\{\lambda _i,\mu
_j\}=\delta _{ij}  \label{2}
\end{equation}
is called separated \cite{sk1} if there exist $n$ relations of the form 
\begin{equation}
\varphi _i(\lambda _i,\mu _i,h_1,...,h_n)=0,\,\,\,\,\,\,i=1,...,n  \label{3}
\end{equation}
joining each pair $(\mu _i,\lambda _i)$ of conjugate coordinates and all
Hamiltonians $h_k,\,k=1,...,n.$ Fixing the values of Hamiltonians $%
h_j=const_j=a_j$ one obtains from (\ref{3}) an explicit factorization of the
Liouville tori given by the equations 
\begin{equation}
\varphi _i(\lambda _i,\mu _i,a_1,...,a_n)=0,\,\,\,\,\,\,i=1,...,n.  \label{4}
\end{equation}
In the Hamilton-Jacobi method, for a given Hamiltonian function $h_r(\mu
,\lambda )$ we are looking for a canonical transformation $(\mu ,\lambda
)\rightarrow (a,b)$ in the form $b_i=\frac{\partial W}{\partial a_i},\mu _i=%
\frac{\partial W}{\partial \lambda _i},$ where $W(\lambda ,a)$ is a
generating function given by the related Hamilton-Jacobi equation 
\begin{equation}
h_r(\lambda ,\frac{\partial W}{\partial \lambda })=a_r.  \label{5}
\end{equation}
If $(\mu ,\lambda )$ are separated coordinates, then 
\begin{equation}
W(\lambda ,a)=\sum_{i=1}^nW_i(\lambda _i,a)  \label{6}
\end{equation}
and the partial differential equation (\ref{5}) splits into $n$ ordinary
differential equations 
\begin{equation}
\varphi _i(\lambda _i,\frac{\partial W}{\partial \lambda _i}%
,a_1,...,a_n)=0,\,\,\,\,\,\,i=1,...,n  \label{7}
\end{equation}
just of the form (\ref{4}). In $(a,b)$ coordinates the flow is trivial 
\begin{equation}
(a_j)_{t_r}=0,\,\,\,\,(b_j)_{t_r}=\delta _{jr}  \label{8}
\end{equation}
and the implicit form of the trajectories $\lambda _i(t_r)$ is the following 
\begin{equation}
b_j(\lambda ,a)=\frac{\partial W}{\partial a_j}=\delta
_{jr}t_r+const,\,\,\,\,j=1,...,n.  \label{9}
\end{equation}
So, given a set of separated variables, it is possible to solve a related
dynamical system by quadratures. In the $19th$ century and most of the
present century, for a number of models of classical mechanics the separated
variables were either guessed or found by some \emph{ad hoc }methods. For
example, in the second half of the $19$\emph{th} century, Neumann's
investigation of a particle moving on a sphere under the action of a linear
force \cite{ne} and Jacobi's study of the geodesics motion on an ellipsoid 
\cite{ja} exploited the separability of the Hamilton-Jacobi equation to
solve the equations of motion by quadratures. In $1891$ St\"{a}ckel
initiated the program dealing the classification of Hamiltonian systems
according to their separability or nonseparability, presenting conditions
for separability of the Hamilton-Jacobi equation in orthogonal coordinates 
\cite{st}. For three dimensional flat space, the $11$ possible coordinate
systems in which separation may take place were deduced in a paper by
Eisenhart \cite{ei1}. They were all obtained as degenerations of the
confocal ellipsoidal coordinates \cite{mil}. For each of the coordinates
Eisenhart \cite{ei2} determined the form of the potential that permitted a
separation of variables. These potentials, designated St\"{a}ckel or
separable potentials, played a crucial role in Hamiltonian mechanics before
the development of more qualitative geometric methods for differential
equations. Although in all classical papers the transformation to separated
coordinates was searched in the form of point transformations, nevertheless
they can be produced by an arbitrary canonical transformation involving both
coordinates and momenta.

A fundamental progress in the theory of separability was made in 1985, when
Sklyanin adopted the method of soliton systems, i.e. the Lax representation,
to systematic derivation of separated variables (see his review article \cite
{sk1}). It was the first constructive theory of separated coordinates for
dynamical systems. In his approach, the appropriate Hamiltonians appear as
coefficients of the spectral curve, i.e. the characteristic equation of Lax
matrix. His method was successfully applied to separation of variables for
many old and new integrable systems \cite{sk2}-\cite{kuz}.

In this paper we present a new constructive separability theory, which has
been recently intensely developed, based on a bi-Hamiltonian property of
integrable systems. In last decade a considerable progress has been made in
construction of new integrable finite dimensional dynamical systems showing
bi-Hamiltonian property. The majority of them originate from stationary
flows, restricted flows or nonlinearization of Lax equations of underlying
soliton systems \cite{an1}-\cite{sr}. Quite recently a fundamental property
of such systems has been discovered, i.e. their separability. It was proved 
\cite{1}-\cite{6} that most bi-Hamiltonian finite dimensional chains, which
start with a Casimir of the first Poisson structure and terminate with a
Casimir of the second Poisson structure, are integrable by quadratures,
through the solutions of the appropriate Hamilton-Jacobi equation.

The presented review article is based on results of papers \cite{1}-\cite{6}
systematizing and unifying them into a compact separability theory in the
frame of the set of canonical coordinates. The results derived so far are
sufficiently promising to consider the theory as an alternative or
complement to the Sklyanin ones.

\section{Preliminaries}

Let us re-examine some facts about bi-Hamiltonian systems. We recall some
definitions. Let $M$ be a differentiable manifold, $TM$ and $T^{*}M$ its
tangent and cotangent bundle. At any point $u\in M,$ the tangent and
cotangent spaces are denoted by $T_uM$ and $T_u^{*}M$, respectively. The
pairing between them is given by the map $<\cdot ,\cdot >:$ $T_u^{*}M\times
T_uM\rightarrow R.$ For each smooth function $F\in C^\infty (M),$ $dF$
denotes the differential of $F$. $M$ is said to be a Poisson manifold if it
is endowed with a Poisson bracket $\{\cdot ,\cdot \}:C^\infty (M)\times $ $%
C^\infty (M)\rightarrow C^\infty (M),$ in general degenerate. The related
Poisson tensor $\pi $ is defined by $\{F,G\}_\pi (u)\,:=<dG,\pi \circ
dF>(u)\,=<dG(u),\pi (u)dF(u)>$. So, at each point $u,\,\pi (u)$ is a linear
map $\pi (u):T_u^{*}M\rightarrow T_uM$ which is skew-symmetric and fulfils
the Jacobi identity. Any function $c\in C^\infty (M),$ such that $dc\in \ker
\pi ,$ is called a Casimir of $\pi .$ Let $\pi _0$,$\pi _1:T^{*}M\rightarrow
TM$ be two Poisson tensors on $M.$ A vector field $K$ is said to be a
bi-Hamiltonian with respect to $\pi _0$ and $\pi _1$ if there exist two
smooth functions $H$,$F\in C^\infty (M)$ such that 
\begin{equation}
K=\pi _0\circ dH=\pi _1\circ dF.  \label{10}
\end{equation}
Poisson tensors $\pi _0$ and $\pi _1$ are said to be compatible if the
associated pencil $\pi _\lambda =\pi _1-\lambda \pi _0$ is itself a Poisson
tensor for any $\lambda \in R.$ Moreover, if $\pi _0$ is invertible, the
tensor $N=\pi _1\circ \pi _0^{-1}$, called a recursion operator, is a
Nijenhuis (hereditary) tensor of such a property that when it acts on a
given bi-Hamiltonian vector field $K$, it produces another bi-Hamiltonian
vector field being a symmetry generator of $K$. Hence, having the invariant
Nijenhuis tensor, one can construct a hierarchy of Hamiltonian symmetries
and related hierarchy of constants of motion for an underlying system, so
important for its integrability.

Unfortunately, for majority of bi-Hamiltonian finite dimensional systems,
both Poisson structures are degenerate, so one cannot construct the
recursion Nijenhuis tensor inverting one of the Poisson structures.
Nevertheless, due to the nonuniqueness of Hamiltonian functions, determined
up to an appropriate Casimir function, it is always possible to construct a
finite bi-Hamiltonian chain starting and terminating with Casimirs of $\pi
_0 $ and $\pi _1$, respectively.

\section{One-Casimir chains}

Let us consider a Poisson manifold $M$ of $\dim M=2n+1$ equipped with a
linear Poisson pencil 
\begin{equation}
\pi _\lambda =\pi _1-\lambda \pi _0  \label{11}
\end{equation}
of maximal rank, where $\pi _0$ and $\pi _1$ are compatible Poisson
structures and $\lambda $ is a continuous parameter. As was first shown by
Gel'fand and Zakharevich \cite{gel}, a Casimir of the pencil is a polynomial
in $\lambda $ of an order $n$ 
\begin{equation}
h_\lambda =h_0\lambda ^n+h_1\lambda ^{n-1}+...+h_n  \label{12}
\end{equation}
and generates a bi-Hamiltonian chain 
\begin{equation}
\pi _\lambda \circ dh_\lambda =0\Longleftrightarrow 
\begin{array}{l}
\pi _0\circ dh_0=0 \\ 
\pi _0\circ dh_1=K_1=\pi _1\circ dh_0 \\ 
\pi _0\circ dh_2=K_2=\pi _1\circ dh_1 \\ 
\,\,\,\,\,\,\,\,\,\,\,\,\,\,\,\,\,\,\,\,\,\,\,\,\,\,\,\,\,\,\,\,\vdots  \\ 
\pi _0\circ dh_n=K_n=\pi _1\circ dh_{n-1} \\ 
\qquad \qquad \,\,\,\,\,\,\,\,\,\,\,0=\pi _1\circ dh_n,\,\,
\end{array}
\label{13}
\end{equation}
where $K\equiv K_1,\,\,H\equiv h_1$and $F\equiv h_0.$ In the paper we
restrict our considerations to a class of canonical coordinates $(q,p,c),$
where $q=(q_1,...,q_n)^T$, $p=(p_1,...,p_n)^T$ are generalized coordinates
and $c$ is a Casimir coordinate. Hence, $\pi _0$ always stays a canonical
Poisson matrix. This restriction simplifies the theory in the sense that it
makes a Marsden-Ratiu projection procedure \cite{mar}, \cite{cas} trivial.

\subsection{Darboux-Nijenhuis representation}

To understand the theory better, let us start from the end in some sense,
i.e. from a system written in separated coordinates $(\mu _i,\lambda
_i)_{i=1}^n.$ An interesting observation is that such a system can be
represented by $n$ different points of some curve. We start from Gel'fand
and Zakharevich case. Actually, let us consider a curve in $(\lambda ,\mu )$
plane, in the particular form 
\begin{equation}
f(\lambda ,\mu )=h_\lambda ,\,\,\,\,\,\,\,\,\,\,\,\,\,\,\,\,\,h_\lambda
=c\lambda ^n+h_1\lambda ^{n-1}+...+h_n,  \label{14}
\end{equation}
where $f(\lambda ,\mu )$ is an arbitrary smooth function. Then, let us take $%
n$ different points $(\mu _i,\lambda _i)$ from the curve: 
\begin{equation}
f(\lambda _i,\mu _i)=c\lambda _i^n+h_1\lambda
_i^{n-1}+...+h_n,\,\,\,\,i=1,...,n,  \label{15}
\end{equation}
which will define our separated coordinates as they are of the form (\ref{3}%
). The explicit dependence of $h_k$ on $(\mu _i,\lambda _i,c)_{i=1}^n$ is
given by the solution of $n$ linear equations (\ref{15}), while for fixed
values of $h_k=a_k$ and $\mu _i=\frac{\partial W_i}{\partial \lambda _i}$
the system (\ref{15}) allows us to solve the appropriate Hamilton-Jacobi
equations. \newline
Here, we have to mention that the idea to relate the multi-Hamiltonian
property to an $m$-parameter family of curves comes from P. Vanhaecke \cite
{van}.$\,$

In refs. \cite{1}-\cite{4} the bi-Hamiltonian chain was constructed for a 
\emph{separation curve }in the form (\ref{14}). Actually, the Hamiltonian
functions $h_k$ found from the system (\ref{15}) take the following compact
form 
\begin{equation}
\begin{split}
& h_k(\lambda ,\mu ,c)=\sum_{i=1}^n\rho _{k-1}^i(\lambda )\frac{f(\lambda
_i,\mu _i)}{\Delta _i(\lambda )}+c\rho _k(\lambda ),\,\,\,k=1,...,n \\
& h_0=c,
\end{split}
\label{16}
\end{equation}
where 
\begin{gather}
\,\,\Delta _i(\lambda ):=\prod_{j\neq i}(\lambda _i-\lambda _j),  \label{17}
\\
\rho _k(\lambda ):=(-1)^k\sum_{%
\mbox{\tiny
$\begin{array}{c}
j_1,..,j_k \\ 
j_1<..<j_k
\end{array}$
}}\lambda _{j_1}\cdot ...\cdot \lambda _{j_k},\,\,\,\,\,\,\,\,k=1,...,n,
\label{18}
\end{gather}
are the so-called Viete polynomials (symmetric polynomials) and 
\begin{equation}
\rho _{k-1}^i(\lambda ):=\rho _{k-1}(\lambda _i=0)=-\frac{\partial \rho
_k(\lambda )}{\partial \lambda _i}.  \label{19}
\end{equation}
For example, for $n=2$ we have 
\begin{equation}
\rho _1=-\lambda _1-\lambda _2,\,\,\,\rho _2=\lambda _1\lambda _2
\label{19a}
\end{equation}
and for $n=3$%
\begin{equation}
\rho _1=-\lambda _1-\lambda _2-\lambda _3,\,\,\,\rho _2=\lambda _1\lambda
_2+\lambda _1\lambda _3+\lambda _2\lambda _3,\,\,\,\rho _3=-\lambda
_1\lambda _2\lambda _3.  \label{19b}
\end{equation}
The bi-Hamiltonian chain (\ref{13}) is constructed with respect to the
following compatible Poisson matrix 
\begin{equation}
\pi _0=\left( 
\begin{array}{rrr}
0 & I & 0 \\ 
-I & 0 & 0 \\ 
0 & 0 & 0
\end{array}
\right) ,\,\,\,\pi _1=\left( 
\begin{array}{rrr}
0\,\,\,\,\,\,\,\,\,\, & \Lambda \,\,\,\,\,\,\,\,\,\,\, & h_{1,\mu } \\ 
-\Lambda \,\,\,\,\,\,\,\,\,\, & 0\,\,\,\,\,\,\,\,\,\,\,\, & -h_{1,\lambda }
\\ 
-\left( h_{1,\mu }\right) ^T & \left( h_{1,\lambda }\right) ^T & 0\,\,\,\,
\end{array}
\right) ,  \label{20}
\end{equation}
where $\Lambda =diag(\lambda _1,...,\lambda _n)$ and $h_{1,\mu }:=\left( 
\frac{\partial h_1}{\partial \mu _1},...,\frac{\partial h_1}{\partial \mu _n}%
\right) ^T.$ Notice that the last column of $\pi _1$ is just the first
vector field $K_1.$ All Hamiltonians $h_k$ are in involution with respect to
both Poisson structures $\pi _0$ and $\pi _1$.

Applying the following important relations 
\begin{gather}
\rho _r(\lambda ) =-\sum_{i=1}^n\rho _{r-1}^i\frac{\lambda _i^n}{\Delta _i},
\label{21} \\
-\sum_{i=1}^n\frac{\partial \rho _r}{\partial \lambda _i}\frac{\lambda _i^m}{%
\Delta _i} =\sum_{i=1}^n\frac{\rho _{r-1}^i(\lambda )\,\lambda _i^m}{\Delta
_i}=\left\{ 
\begin{array}{l}
1,\,\,\,\,\,\,\,\,\,\,m=n-r \\ 
0,\,\,\,\,\,\,\,\,\,\,m\neq n-r
\end{array}
\right\} ,r=1,..,n  \label{21a}
\end{gather}
and the decomposition (\ref{6}), the Hamilton-Jacobi equations 
\begin{equation}
h_r(\lambda ,\frac{\partial W}{\partial \lambda },c)=a_r,\,\,\,\,r=1,...,n
\label{22}
\end{equation}
turn into the form 
\begin{equation}
\sum_{k=1}^n\frac{\rho _{r-1}^k(\lambda )[f(\lambda _k,\partial W_k/\partial
\lambda _k)-c\lambda _k^n]}{\Delta _k}=a_r,\,\,\,r=1,...,n  \label{23}
\end{equation}
with the solution 
\begin{equation}
f(\lambda _k,\partial W_k/\partial \lambda _k,c)=c\lambda _k^n+a_1\lambda
_k^{n-1}+...+a_n,\,\,\,\,k=1,...,n.  \label{24}
\end{equation}
Hence, $W(\lambda ,a)$ can be obtained by solving $n$ decoupled first-order
ODEs (\ref{24}) and the family of dynamical systems (\ref{13}) can be solved
by quadratures.

Now, let us pass to the projection of the Poisson pencil $\pi _\lambda $
onto a symplectic leaf $S$ of $\pi _0$ ($\dim S=2n$) fixing the value of $c$%
. Generally, one has to apply the Marsden-Ratiu theorem which in this case
is trivial, as obviously $\theta _\lambda =\theta _1-\lambda \theta _0,$
where 
\begin{equation}
\theta _0=\left( 
\begin{array}{cc}
0 & I \\ 
-I & 0
\end{array}
\right) ,\,\,\,\,\,\theta _1=\left( 
\begin{array}{cc}
0 & \Lambda \\ 
-\Lambda & 0
\end{array}
\right) ,  \label{25}
\end{equation}
is a nondegenerate Poisson pencil on $S$. Hence, the related Nijenhuis
tensor 
\begin{equation}
N=\theta _1\circ \theta _0^{-1}=\left( 
\begin{array}{cc}
\Lambda & 0 \\ 
0 & \Lambda
\end{array}
\right)  \label{26}
\end{equation}
is diagonal and this is the reason why we will refer below to the separated
coordinates as to the Darboux-Nijenhuis (DN) coordinates. Notice that $\rho
_i$ (\ref{18}) are coefficients of minimal polynomial of the Nijenhuis
tensor 
\begin{equation}
(\det (N-\lambda ))^{1/2}=\lambda ^n+\sum_{i=1}^n\rho _i\lambda
^{n-i}=\prod_{i=1}^n(\lambda -\lambda _i).  \label{26a}
\end{equation}
On $S$ the chain (\ref{13}) turns into the form 
\begin{equation}
\begin{split}
\theta _0\circ dh_1={}& \overline{K}_1=-\frac 1{\rho _n}\theta _1\circ dh_n
\\
\theta _0\circ dh_2={}& \overline{K}_2=-\frac{\rho _1}{\rho _n}\theta
_1\circ dh_n+\theta _1\circ dh_1 \\
& \vdots \\
\theta _0\circ dh_n={}& \overline{K}_n=-\frac{\rho _{n-1}}{\rho _n}\theta
_1\circ dh_n+\theta _1\circ dh_{n-1}.
\end{split}
\label{27}
\end{equation}
The last equation terminates the sequence of vector fields $\overline{K}_r$
in the hierarchy as for the next equation from the chain we have 
\begin{equation}
\theta _0\circ dh_{n+1}=\overline{K}_{n+1}=-\theta _1\circ dh_n+\theta
_1\circ dh_n=0.  \label{28}
\end{equation}
Obviously $N$ is not a recursion operator for the hierarchy (\ref{27}).
Because of the form of the first equation, the vector field $\overline{K}_1$
is called a quasi-bi-Hamiltonian \cite{bro}-\cite{ton} and the chain (\ref
{27}) could be treated alternatively as a starting point of the separability
theory for the case of $c=0$.

\subsection{Arbitrary canonical representation}

Now, let us consider an arbitrary canonical transformation 
\begin{equation}
(q,p)\rightarrow (\lambda ,\mu )  \label{29}
\end{equation}
independent of a Casimir coordinate $c$ (not necessarily a point
transformation!). The advantage of staying inside such a class of
transformations is that the clear structure of a pencil is preserved and the
Marsden-Ratiu projection of the Poisson pencil is still trivial. Of course,
the most general case of multi-Hamiltonian separability theory takes place
when one goes beyond the set of canonical coordinates. i.e. when one tries
to find DN coordinates starting from the pencil written in a non-canonical
representation. But then the simple structure of degenerated Poisson pencil
is lost and the nontrivial problem of the Marsden-Ratiu projection for such
pencil appears (see for example ref. \cite{fal}).

Applying the transformation (\ref{29}) to Hamiltonian functions (\ref{16})
and Poisson matrices (\ref{20}) one finds that 
\begin{equation}
h_k(q,p,c)=h_k(q,p)+cb_k(q,p),\,\,\,k=1,...,n  \label{30}
\end{equation}
and 
\begin{equation}  \label{31}
\begin{split}
\pi _0 &=\left( 
\begin{array}{cc}
\theta _0 & 0 \\ 
0 & 0
\end{array}
\right) ,\,\,\,\theta _0=\left( 
\begin{array}{cc}
0 & I \\ 
-I & 0
\end{array}
\right) , \\
\pi _1 &=\left( 
\begin{array}{cc}
\theta _1 & \overline{K}_1 \\ 
-\overline{K}_1^T & 0
\end{array}
\right) ,\,\,\,\theta _1=\left( 
\begin{array}{cc}
D(q,p) & A(q,p) \\ 
-A^T(q,p) & B(q,p)
\end{array}
\right) ,
\end{split}
\end{equation}
where $A,B$ and $D$ are $n\times n$ matrices. The nondegenerate Poisson
pencil $\theta _\lambda $ on $S$ gives rise to the related Nijenhuis tensor $%
N$ and its adjoint $N^{*}$ in $(q,p)$ coordinates in the form 
\begin{equation}
N=\theta _1\circ \theta _0^{-1}=\left( 
\begin{array}{cc}
A & -D \\ 
B & A^T
\end{array}
\right) ,\,\,\,N^{*}=\theta _0^{-1}\circ \theta _1=\left( 
\begin{array}{cc}
A^T & -B \\ 
D & A
\end{array}
\right) .  \label{33}
\end{equation}

Obviously, in a real situation we start from a given bi-Hamiltonian chain (%
\ref{30})-(\ref{33}) in canonical coordinates $(q,p,c),$ derived by some
method (see for example \cite{an1}-\cite{sr}), and trying to find the DN
coordinates which diagonalize the appropriate Nijenhuis tensor and are
separated coordinates for the considered system. So now we pass to a
systematic derivation of the inverse of transformation (\ref{29}).

The important intermediate step of the construction of DN coordinates are
the so called Hankel-Fr\"{o}benious (HF) non-canonical coordinates $(u,v)$ 
\cite{van}, \cite{mag1}, related to the DN ones through the following
transformation 
\begin{equation}  \label{34}
\begin{split}
u_i &=\rho _i(\lambda _1,...,\lambda _n), \\
\mu _i &=\sum_{k=1}^nv_k\lambda _i^{n-k},\,\,\,\,\,i=1,...,n.
\end{split}
\end{equation}
In $(u,v)$ coordinates one finds 
\begin{gather*}
\theta _0 =\left( 
\begin{array}{cc}
0 & U \\ 
-U^T & 0
\end{array}
\right) ,\,\,\,U=\left( 
\begin{array}{cccc}
0 & 0 & \cdots & 1 \\ 
0 & \cdots & 1 & u_1 \\ 
\cdots & \cdots & \cdots & \cdots \\ 
1 & u_1 & \cdots & u_{n-1}
\end{array}
\right) , \\
N =\left( 
\begin{array}{cc}
F & 0 \\ 
0 & F
\end{array}
\right) ,\,\,\,\,F=\left( 
\begin{array}{ccccc}
-u_1 & 1 & \cdots & \cdots & 0 \\ 
-u_2 & 0 & 1 & \cdots & 0 \\ 
\cdots & \cdots & \cdots & \cdots & \cdots \\ 
-u_{n-1} & 0 & \cdots & \cdots & 1 \\ 
-u_n & 0 & \cdots & \cdots & 0
\end{array}
\right) , \\
\theta _1 = N\circ \theta _0=\left( 
\begin{array}{cc}
0 & FU \\ 
-FU^T & 0
\end{array}
\right) ,\,\,\,\,\,N^{*}=N^T.
\end{gather*}
Moreover the differentials $du_i,\,dv_i$ satisfy the following recursion
relations \cite{mag1}, \cite{fal} 
\begin{equation}  \label{35}
\begin{array}{@{}lcl}
N^{*}\circ du_1 & = & du_2-u_1du_1, \\ 
N^{*}\circ du_2 & = & du_3-u_2du_1, \\ 
& \vdots &  \\ 
N^{*}\circ du_{n-1} & = & du_n-u_{n-1}du_1, \\ 
N^{*}\circ du_n & = & -u_ndu_1, \\[1em] 
N^{*}\circ dv_1 & = & dv_2-u_1dv_1, \\ 
N^{*}\circ dv_2 & = & dv_3-u_2dv_1, \\ 
& \vdots &  \\ 
N^{*}\circ dv_{n-1} & = & dv_n-u_{n-1}dv_1, \\ 
N^{*}\circ dv_n & = & -u_ndv_1.
\end{array}
\end{equation}
Note that vector fields $\theta _0\circ du_1$ and $\theta _0\circ dv_1$ are
quasi-bi-Hamiltonian.

Now we relate the canonical coordinates $(q,p,c)$ to the DN separated
coordinates $(\lambda ,\mu ,c)$. From the minimal polynomial of $N\,$ (\ref
{33}) we get 
\begin{equation}
u_k=\zeta _k(q,p),\,\,\,k=1,...,n.  \label{36}
\end{equation}
Conjugate coordinates $v_k=v_k(q,p),k=2,...,n$ are found from the recursion
formula (\ref{35}) while $v_1=v_1(q,p)$ coordinate from relations 
\begin{equation}
\{u_j,v_1\}_{\theta _k}=\delta
_{j,n-k},\,\,\,\,\,\,\,\,\,\,j=1,...,n,\,\,\,\,\,\,k=0,...n-1.  \label{37}
\end{equation}
Hence we get 
\begin{equation}
v_k=\vartheta _k(q,p),\,\,\,k=1,...,n.  \label{38}
\end{equation}
Eliminating $(u,v)$ coordinates from (\ref{34}), (\ref{36}) and (\ref{38})
we derive the desired relations 
\begin{equation}
\chi _i(q,p;\lambda ,\mu )=0,\,\,\,\,\,i=1,...,2n.  \label{39}
\end{equation}

Now let us concentrate on a special but important case of point
transformation between $(q,p,c)$ and $(\lambda ,\mu ,c)$ variables. Then 
\begin{equation}  \label{40a}
\begin{split}
\theta _1 &=\left( 
\begin{array}{cc}
0 & A(q) \\ 
-A^T(q) & B(q,p)
\end{array}
\right) , \\
N &=\left( 
\begin{array}{cc}
A(q) & 0 \\ 
B(q,p) & A^T(q)
\end{array}
\right) ,
\end{split}
\end{equation}
where matrix elements of $B$ are at most linear in $p$ coordinates, so
coefficients of minimal polynomial of $N$ are equal to coefficients of the
characteristic polynomial of $A$. Hence, we find the first part of the
canonical transformation in the form 
\begin{equation}
\rho _i(\lambda )=\eta _i(q),\,\,\,i=1,...,n\,\,\Longrightarrow \,\,q_k=\psi
_k(\lambda ),\,\,\,k=1,...,n\,.  \label{41}
\end{equation}
The complementary part of the transformation we get from the generating
function 
\begin{equation}
G(p,\lambda )=\sum_{i=1}^np_i\psi _i(\lambda ).  \label{42}
\end{equation}
Then, 
\begin{equation}
\mu _i=\frac{\partial G}{\partial \lambda _i},\,\,i=1,...,n\,\,\,%
\Longrightarrow \,\,p_k=\varphi _k(\lambda ,\mu ),\,\,\,k=1,...,n.
\label{43}
\end{equation}

At the end of this subsection we introduce the notion of an inverse
bi-Hamiltonian separable chains. In ref. \cite{4} it was demonstrated that
for each separable bi-Hamiltonian chain (\ref{13}), (\ref{30}), (\ref{31}), (%
\ref{33}) in canonical coordinates $(q,p,c)$ there exists a related inverse
bi-Hamiltonian separable chain 
\begin{equation}
\begin{array}{@{}l@{}c@{}c@{}l}
\pi _0\circ dh_{n+1}^{\prime } & {}={} & 0 &  \\ 
\pi _0\circ dh_n^{\prime } & {}={} & K_n^{\prime } & =\pi _{-1}\circ
dh_{n+1}^{\prime } \\ 
&  & \vdots &  \\ 
\pi _0\circ dh_r^{\prime } & {}={} & K_r^{\prime } & =\pi _{-1}\circ
dh_{r+1}^{\prime } \\ 
&  & \vdots &  \\ 
\pi _0\circ dh_1^{\prime } & {}={} & K_1^{\prime } & =\pi _{-1}\circ
dh_2^{\prime } \\ 
&  & 0 & =\pi _{-1}\circ dh_1^{\prime }
\end{array}
\label{44}
\end{equation}
where $h_{n+1}^{\prime }=h_0=c,$ 
\begin{equation}
\pi _{-1}=\left( 
\begin{array}{cc}
\theta _{-1} & \overline{K}_n^{\prime } \\ 
-\overline{K}_n^{\prime T} & 0
\end{array}
\right) ,\,\,\,\,\theta _{-1}=\theta _0\circ \theta _1^{-1}\circ \theta
_0=N^{-2}\circ \theta _1,  \label{45}
\end{equation}
$\overline{K}_n^{\prime }=\theta _0\circ dh_n^{\prime }$ and 
\begin{equation}
h_r^{\prime }(q,p,c)=h_r(q,p)+cb_r^{\prime }(q),\,\,\,\,\,\,b_r^{\prime }(q)=%
\frac{b_{r-1}(q)}{b_n(q)},\,\,\,\,r=1,...,n.  \label{46}
\end{equation}
Notice that in both chains the respective Hamiltonians (\ref{30}), (\ref{46}%
) and related vector fields differ only by the $c-$dependent parts. In the
case of a point transformation to the DN coordinates, when $\theta _1$ takes
the form (\ref{40a}), we get 
\begin{equation}
\theta _1^{-1}=\left( 
\begin{array}{cc}
\left( A^{-1}\right) ^T\circ D\circ A^{-1} & -\left( A^{-1}\right) ^T \\ 
A^{-1} & 0
\end{array}
\right)  \label{47}
\end{equation}
and 
\begin{equation}
\theta _0\circ \theta _1^{-1}\circ \theta _0=\left( 
\begin{array}{cc}
0 & A^{-1} \\ 
-\left( A^{-1}\right) ^T & -\left( A^{-1}\right) ^T\circ D\circ A^{-1}
\end{array}
\right) .  \label{48}
\end{equation}

Notice that both chains (\ref{13}) and (\ref{44}) given in natural
coordinates can be transformed to the Nijenhuis bi-Hamiltonian form (\ref{16}%
)-(\ref{20}), where the canonical transformation can be derived from the
relation $\rho _r(\lambda )=b_r(q),\,\,r=1,...,n$ in the first case and from
the relations $\rho _r(\lambda )=\frac{b_{n-r}(q)}{b_n(q)},\,r=1,...,n$ in
the second case.

Consider the set of extended functions 
\begin{equation}
h_r(q,p;c,c^{\prime })=h_r(q,p)+cb_r(q)+c^{\prime }b^{\prime }(q).
\label{49}
\end{equation}
They can be simultaneously put into the bi-Hamiltonian and the inverse
bi-Hamiltonian hierarchies. In the first case $c$ is treated as the Casimir
variable and $c^{\prime }$ as the parameter and in the second case $c$ is
treated as the parameter and $c^{\prime }$ as the Casimir variable.

\subsection{Examples}

We shall illustrate the theory presented by a few representative examples.
More examples can be found in refs. \cite{1}-\cite{4}. In all examples from
this section canonical transformations between natural and separated
coordinates are point like. An example of a nonpoint transformation will be
given at the end of this paper.

\begin{example}
\emph{The one-Casimir extension of the Henon-Heiles system.} \newline
Let us consider the integrable case of the Henon-Heiles system generated by
the Hamiltonian $H=\frac 12p_1^2+\frac 12p_2^2+q_1^3+\frac 12q_1q_2^2.$ Its
one-Casimir extension reads \cite{bl5} 
\begin{equation}
(q_1)_{tt}=-3q_1^2-\frac 12q_2^2+c,\,\,\,\,(q_2)_{tt}=-q_1q_2  \label{50}
\end{equation}
and belongs to the bi-Hamiltonian chain 
\begin{equation}
\begin{array}{l}
\pi _0\circ dh_0=0 \\ 
\pi _0\circ dh_1=K_1=\pi _1\circ dh_0 \\ 
\pi _0\circ dh_2=K_2=\pi _1\circ dh_1 \\ 
\qquad \qquad \,\,\,\,\,\,0\,\,\,\,=\pi _1\circ dh_2,\,\,
\end{array}
\label{51}
\end{equation}
where

\begin{equation}
\begin{split}
h_0& =c, \\
h_1& =h_1(q,p)+cb_1(q)=\frac 12p_1^2+\frac 12p_2^2+q_1^3+\frac
12q_1q_2^2-cq_1, \\
h_2& =h_2(q,p)+cb_2(q)=\frac 12q_2p_1p_2-\frac 12q_1p_2^2+\frac
1{16}q_2^4+\frac 14q_1^2q_2^2-\frac 14cq_2^2, \\
&
\,\,\,\,\,\,\,\,\,\,\,\,\,\,\,\,\,\,\,\,\,\,\,\,\,\,\,\,\,\,\,\,\,\,\,\,\,\,%
\,\,\,\,\,\,\,\,\,\,\,\,\,\,\,\,\,\,{}
\end{split}
\label{52}
\end{equation}
\begin{equation}
\pi _1=\left( 
\begin{array}{ccccc}
0 & 0 & q_1 & \frac 12q_2 & p_1 \\ 
0 & 0 & \frac 12q_2 & 0 & p_2 \\ 
-q_1 & -\frac 12q_2 & 0 & \frac 12p_2 & -h_{1,q_1} \\ 
-\frac 12q_2 & 0 & -\frac 12p_2 & 0 & -h_{1,q_2} \\ 
-p_1 & -p_2 & h_{1,q_1} & h_{1,q_2} & 0
\end{array}
\right) .  \label{53}
\end{equation}
The construction of the related inverse bi-Hamiltonian chain, according to
the results from the previous subsection, gives the following results: 
\begin{equation}
\begin{array}{@{}l@{\ }l@{\ }l}
\pi _0\circ dh_3^{\prime }= & 0 &  \\ 
\pi _0\circ dh_2^{\prime }= & K_2^{\prime } & =\pi _{-1}\circ dh_3^{\prime }
\\ 
\pi _0\circ dh_1^{\prime }= & K_1^{\prime } & =\pi _{-1}\circ dh_2^{\prime }
\\ 
& 0 & =\pi _{-1}\circ dh_1^{\prime },
\end{array}
\label{54}
\end{equation}
where 
\begin{equation}
\begin{split}
h_1^{\prime }& =h_1(q,p)-\frac 4{q_2^2}c, \\
h_2^{\prime }& =h_2(q,p)+\frac{4q_1}{q_2^2}c, \\
h_3^{\prime }& =c,
\end{split}
\label{55}
\end{equation}
\begin{equation}
\pi _{-1}=\left( 
\begin{array}{ccccc}
0 & 0 & 0 & 2/q_2 & \frac 12q_2p_2 \\ 
0 & 0 & 2/q_2 & -4q_1/q_2^2 & \frac 12q_2p_1-q_1p_2 \\ 
0 & -2/q_2 & 0 & 2p_2/q_2^2 & -h_{2,q_1}^{\prime } \\ 
-2/q_2 & 4q_1/q_2^2 & -2p_2/q_2^2 & 0 & -h_{2,q_2}^{\prime } \\ 
-\frac 12q_2p_2 & -\frac 12q_2p_1+q_1p_2 & h_{2,q_1}^{\prime } & 
h_{2,q_2}^{\prime } & 0
\end{array}
\right)   \label{56}
\end{equation}
and Newton equations related with the natural Hamiltonian $h_1^{\prime }$
are: 
\begin{equation}
(q_1)_{tt}=-3q_1^2-\frac 12q_2^2,\,\,\,\,(q_2)_{tt}=-q_1q_2-\left( \frac
2{q_2}\right) ^3c,  \label{57}
\end{equation}
being just the second well known one-Casimir extension \cite{an4} of the
Henon-Heiles system considered. Both systems (\ref{50}) and (\ref{57}) can
be transformed to the Nijenhuis chain (\ref{16})--(\ref{20}) through the
respective transformations 
\begin{equation}
\begin{split}
& q_1=\lambda _1+\lambda _2,\,\,\,p_1=\frac{\lambda _1\mu _1}{\lambda
_1-\lambda _2}+\frac{\lambda _2\mu _2}{\lambda _2-\lambda _1}, \\
& q_2=2\sqrt{-\lambda _1\lambda _2},\,\,\,p_2=\sqrt{-\lambda _1\lambda _2}%
\left( \frac{\mu _1}{\lambda _1-\lambda _2}+\frac{\mu _2}{\lambda _2-\lambda
_1}\right) , \\
& f(\lambda _i,\mu _i)=\frac 12\lambda _i\mu _i^2+\lambda _i^4,
\end{split}
\label{58a}
\end{equation}
and 
\begin{equation}
\begin{split}
& q_1=\frac 1{\lambda _1}+\frac 1{\lambda _2},\,\,\,p_1=\lambda _1\lambda
_2\left( \frac{\lambda _1\mu _1}{\lambda _1-\lambda _2}+\frac{\lambda _2\mu
_2}{\lambda _2-\lambda _1}\right) , \\
& q_2=\frac 2{\sqrt{-\lambda _1\lambda _2}},\,\,\,p_2=-\sqrt{-\lambda
_1\lambda _2}\left( \frac{\lambda _1^2\mu _1}{\lambda _1-\lambda _2}+\frac{%
\lambda _2^2\mu _2}{\lambda _2-\lambda _1}\right) , \\
& f(\lambda _i,\mu _i)=\frac 12\lambda _i^4\mu _i^2+\lambda _i^{-3}.
\end{split}
\label{59a}
\end{equation}
\end{example}

\begin{example}
\emph{One-Casimir extension of the Kepler problem in the plane.}\newline
Let us consider the classical problem of a particle in the plane under the
influence of the Kepler potential and an additional homogeneous field force.
The Hamiltonian function reads 
\begin{equation}
h_1(q,p,c)=\frac 12p_1^2+\frac 12p_2^2-\frac a{\sqrt{q_1^2+q_2^2}%
}-cq_2,\,\,\,\,\,a=const.  \label{60}
\end{equation}
There is a second independent integral of the motion 
\begin{equation}
h_2(q,p,c)=-\frac 12q_2p_1^2+\frac 12q_1p_1p_2+\frac 12\frac{aq_2}{\sqrt{%
q_1^2+q_2^2}}-\frac 14cq_1^2,  \label{61}
\end{equation}
which together with $h_0=c$ allows us to construct a bi-Hamiltonian chain (%
\ref{51}) with the second Poisson structure in the form 
\begin{equation}
\pi _1=\left( 
\begin{array}{ccccc}
0 & 0 & 0 & \frac 12q_1 & p_1 \\ 
0 & 0 & \frac 12q_1 & q_2 & p_2 \\ 
0 & -\frac 12q_1 & 0 & -\frac 12p_1 & -h_{1,q_1} \\ 
-\frac 12q_1 & -q_2 & \frac 12p_1 & 0 & -h_{1,q_2} \\ 
-p_1 & -p_2 & h_{1,q_1} & h_{1,q_2} & 0
\end{array}
\right) .  \label{62}
\end{equation}
The inverse bi-Hamiltonian chain (\ref{54}) is given for functions 
\begin{equation}
\begin{split}
& h_1^{\prime }(q,p,c)=\frac 12p_1^2+\frac 12p_2^2-\frac a{\sqrt{q_1^2+q_2^2}%
}-\frac 4{q_1^2}c, \\
& h_2^{\prime }(q,p,c)=-\frac 12q_2p_1^2+\frac 12q_1p_1p_2+\frac 12\frac{aq_2%
}{\sqrt{q_1^2+q_2^2}}+\frac{4q_2}{q_1^2}c, \\
& h_3^{\prime }=c,
\end{split}
\label{63}
\end{equation}
and the second Poisson tensor in the form 
\begin{equation}
\pi _{-1}=\left( 
\begin{array}{ccccc}
0 & 0 & -4q_2/q_1^2 & 2/q_1 & \frac 12q_2p_1-q_1p_2 \\ 
0 & 0 & 2/q_1 & 0 & \frac 12q_2p_2 \\ 
4q_2/q_1^2 & -2/q_1 & 0 & -2p_1/q_1^2 & -h_{2,q_1}^{\prime } \\ 
-2/q_1 & 0 & 2p_1/q_1^2 & 0 & -h_{2,q_2}^{\prime } \\ 
-\frac 12q_2p_1+q_1p_2 & -\frac 12q_2p_2 & h_{2,q_1}^{\prime } & 
h_{2,q_2}^{\prime } & 0
\end{array}
\right) .  \label{64}
\end{equation}
The transformations to DN coordinates for both chains are the following 
\begin{equation}
\begin{split}
& q_1=2\sqrt{\lambda _1\lambda _2},\,\,\,p_1=\sqrt{\lambda _1\lambda _2}%
\left( \frac{\mu _1}{\lambda _1-\lambda _2}+\frac{\mu _2}{\lambda _2-\lambda
_1}\right) , \\
& q_2=\lambda _1+\lambda _2,\,\,\,p_2=\frac{\lambda _1\mu _1}{\lambda
_1-\lambda _2}+\frac{\lambda _2\mu _2}{\lambda _2-\lambda _1}, \\
& f(\lambda _i,\mu _i)=\frac 12\lambda _i\mu _i^2+\frac 12a,
\end{split}
\label{65}
\end{equation}
and 
\begin{equation}
\begin{split}
& q_1=\frac 2{\sqrt{-\lambda _1\lambda _2}},\,\,\,p_1=\sqrt{-\lambda
_1\lambda _2}\left( \frac{\lambda _1^2\mu _1}{\lambda _1-\lambda _2}+\frac{%
\lambda _2^2\mu _2}{\lambda _2-\lambda _1}\right) , \\
& q_2=\frac{\lambda _1+\lambda _2}{\lambda _1\lambda _2},\,\,\,p_2=\lambda
_1\lambda _2\left( \frac{\lambda _1\mu _1}{\lambda _1-\lambda _2}+\frac{%
\lambda _2\mu _2}{\lambda _2-\lambda _1}\right) , \\
& f(\lambda _i,\mu _i)=\frac 12\lambda _i^4\mu _i^2+\frac 12a\lambda _i.
\end{split}
\label{66}
\end{equation}
\end{example}

\begin{example}
\emph{One-Casimir extension of elliptic separable potentials. }\newline
In refs.\cite{1},\cite{3} it was proved that every natural Hamiltonian
system 
\begin{equation}
H=\frac 12\sum_{i=1}^np_i^2+V(q)+\frac 12c(q,q),  \label{67}
\end{equation}
where $(.,.)$ means the scalar product, which admits in extended phase space 
$M\ni (q,p,c)$ the bi-Hamiltonian formulation 
\begin{equation}
\left( 
\begin{array}{c}
q \\ 
p \\ 
c
\end{array}
\right) _t=\pi _0\circ dh_1=\pi _1\circ dh_0,  \label{68}
\end{equation}
where $\pi _0$ is a canonical Poisson structure, 
\begin{equation}
\pi _1=\left( 
\begin{array}{ccc}
0 & A-\frac 12q\otimes q & h_{1,p} \\[2pt] 
-A+\frac 12q\otimes q & \frac 12p\otimes q-\frac 12q\otimes p & -h_{1,q} \\%
[2pt] 
-\left( h_{1,p}\right) ^T & \left( h_{1,q}\right) ^T & 0
\end{array}
\right) ,  \label{69}
\end{equation}
$A=diag(\alpha _1,...,\alpha _N),\,\,\alpha _i-$different positive
constants, $h_0=c,\,h_1=H+c\rho _1(\alpha ),$ is separable in generalized
elliptic coordinates. This bi-Hamiltonian formulation generates the chain (%
\ref{2}) \cite{sr} of commuting bi-Hamiltonian vector fields, where 
\begin{equation}
\begin{split}
& h_r(q,p,c;\alpha )=h_r(q,p;\alpha )+cb_r(q;\alpha ), \\
& b_r(q;\alpha )=\rho _r(\alpha )+\frac 12\sum_{k=1}^{r-1}\rho _k(\alpha
)(q,A^{r-k-1}q),\,\,\,r=1,...,n-1, \\
& b_n(\alpha )=\rho _n(\alpha )[1-\frac 12(q,A^{-1}q)], \\
& h_r(q,p;\alpha )=\sum_{k=0}^r\rho _k(\alpha )\overline{h}_{r-k},\,\,\,%
\overline{h}_s=\frac 12\sum_{i=1}^n\alpha _i^{s-1}R_i, \\
& R_i=\sum_{i\neq j}\frac{q_ip_j-q_jp_i}{\alpha _i-\alpha _j}%
+p_i^2+V_i(q),\,\,\,\sum_{i=1}^nV_i(q)=V(q),
\end{split}
\label{70}
\end{equation}
$\{H,R_i\}_{\pi _0}=0,\,\,i=1,...,n$ and $\rho _r(\alpha )$ are Viete
polynomials of $\alpha .$

On the other hand, according to our procedure, the inverse bi-Hamiltonian
chain (\ref{54}) for one-Casimir extension of potentials separable in
elliptic coordinates reads 
\begin{gather}
h_r^{\prime }(q,p,c;\alpha )=h_r(q,p;\alpha )+c\frac{b_{r-1}(q;\alpha )}{%
b_n(q;\alpha )},  \label{71} \\
\pi _{-1}=\left( 
\begin{array}{ccc}
0 & B^{-1} & h_{n,p} \\[2pt] 
-B^{-1} & -B^{-1}(\frac 12p\otimes q-\frac 12q\otimes p)B^{-1} & -h_{n,q} \\%
[2pt] 
-\left( h_{n,p}\right) ^T & \left( h_{n,p}\right) ^T & 0
\end{array}
\right) ,  \label{72}
\end{gather}
where 
\begin{equation}
\begin{split}
& B=A-\frac 12q\otimes q, \\
& B^{-1}=\frac 1{\left| B\right| }\left[ \partial _\alpha \left| B\right|
+\frac 12(\partial _\alpha q\otimes \partial _\alpha q)\left| A\right|
\right] , \\
& \partial _\alpha =diag\left( \frac \partial {\partial \alpha _1},...,\frac
\partial {\partial \alpha _n}\right) , \\
& \left| B\right| =\left| A\right| -\frac 12(q,\partial _\alpha \left|
A\right| q).
\end{split}
\label{74}
\end{equation}
Notice that now, the natural Hamiltonian in the inverse hierarchy is the
last one of the form 
\begin{equation}
H^{\prime }=h_1^{\prime }(q,p,c)=\frac 12\sum_{i=1}^np_i^2+V(q)+\frac c{\rho
_n(\alpha )[1-(q,A^{-1}q)]}.  \label{75}
\end{equation}
The basic example here is the Garnier system with the potential $V(q)=\frac
14(q,q)^2-\frac 12(q,Aq).$ This potential is a member of an infinite family
of permutationally symmetric potentials separable in generalized elliptic
coordinates \cite{an5}. The point transformation to DN separated coordinates
can be constructed from the relations 
\begin{equation}
\rho _r(\lambda )=\rho _r(\alpha )+\frac 12\sum_{k=1}^{r-1}\rho _k(\alpha
)(q,A^{r-k-1}q),\,\,\,r=1,...,n.  \label{76}
\end{equation}
But so defined DN coordinates are just the generalized elliptic coordinates $%
\lambda _1,...,\lambda _n$ defined by the relation 
\begin{equation}
1+\frac 12\sum_{k=1}^n\frac{q_k^2}{z-\alpha _k}=\frac{\prod_{j=1}^n(z-%
\lambda _j)}{\prod_{k=1}^n(z-\alpha _k)}.  \label{77}
\end{equation}
The proof is given in refs. \cite{1} and \cite{3}.
\end{example}

More examples of separated systems by the method presented the reader can
find in refs. \cite{1}-\cite{4},\cite{mor1}-\cite{ton}.

\section{Multi-Casimir unsplit chains}

In the previous section a separability theory of one-Casimir bi-Hamiltonian
chains was reviewed. Here we pass to the generalization of the theory and
include multi-Casimir cases. This procedure considerably extends the class
of separable systems and in general covers a new class of chains, i.e. the
so-called split chains. In the following section we consider the simplest
generalization of unsplit multi-Casimir chains related to the extension of
the separation curve (\ref{14}) of the form 
\begin{equation}
f(\lambda ,\mu )=h_\lambda ,\,\,\,\,h_\lambda =c_n\lambda
^{2n-1}+...+c_1\lambda ^n+h_1\lambda ^{n-1}+...+h_n.  \label{78}
\end{equation}
Choosing other admissible forms of the separation curve with more then one
Casimir, one can construct split bi(multi)-Hamiltonian chain, i.e. the chain
which splits onto a few bi(multi)-Hamiltonian sub-chains, each starting and
terminating with some Casimir of the appropriate Poisson structure. The work
on split cases is still in progress but some results are presented in the
next section.

\subsection{Multi-Hamiltonian Darboux-Nijenhuis chains}

In ref. \cite{5} the multi-Hamiltonian chain was constructed for the \emph{%
separation curve }in the form (\ref{78}). Actually, the Hamiltonian
functions $h_k$ found from the system 
\begin{equation}
f(\lambda _i,\mu _i)=c_n\lambda _i^{2n-1}+...+c_1\lambda _i^n+h_1\lambda
_i^{n-1}+...+h_n,\,\,\,\,i=1,...,n,  \label{79}
\end{equation}
take the following form 
\begin{equation}
h_r(\lambda ,\mu ,c)=-\sum_{i=1}^n\frac{\partial \rho _r}{\partial \lambda _i%
}\frac{f(\lambda _i,\mu _i)}{\Delta _i}+\sum_{j=1}^nc_j\beta _{j,r}(\lambda
),\,\,\,\,r=1,...,n,  \label{80}
\end{equation}
where $\beta _{1,r}(\lambda )\equiv \rho _r(\lambda )$ and $\beta
_{m,r}\,,m=2,...,n,$ are defined by the recursive formula 
\begin{equation}
\beta _{m,r}=\beta _{m-1,r+1}-\beta _{m-1,1}\cdot \beta _{1,r}.  \label{81}
\end{equation}
Notice that Hamiltonians (\ref{80}) are just Hamiltonians (\ref{16})
supplemented with extra terms, linear with respect to additional Casimirs $%
c_i,\,i=2,...,n.$ On the extended phase space $M\ni (\lambda _1,...,\lambda
_n,\mu _1,...,\mu _n,$ $c_1,...,c_n)$ functions (\ref{80}) and $%
h_{1-r}(c)=c_r,\,\,r=1,...,n$ form $\left( 
\begin{array}{c}
n+1 \\ 
2
\end{array}
\right) $bi-Hamiltonian chains, each generated by a Poisson pencil $\pi
_{\lambda ^{k-1}}=\pi _k-\lambda ^{k-i}\pi _i$ of order $1\leq (k-i)\leq n$, 
\begin{equation}
\pi _{\lambda ^{k-i}}\circ dh_{\lambda \,\,\Longrightarrow \,\,} 
\begin{array}{l@{\ }c@{\ }l@{\ }l}
\pi _i\circ dh_{-i} & = & 0 &  \\ 
\pi _i\circ dh_{-i+1} & = & K_1 & =\pi _k\circ dh_{-k+1} \\ 
\pi _i\circ dh_{-i+2} & = & K_2 & =\pi _k\circ dh_{-k+2} \\ 
&  & \vdots &  \\ 
\pi _i\circ dh_{-i+j} & = & K_j & =\pi _k\circ dh_{-k+j} \\ 
&  & \vdots &  \\ 
\pi _i\circ dh_{-i+n} & = & K_n & =\pi _k\circ dh_{-k+n} \\ 
&  & 0 & =\pi _k\circ dh_{-k+n+1}
\end{array}
\,\,\,\,\,\,\,\,\,\,\,\,\,\,\,\,\,\,\,\,\,\,\,0\leq i<k\leq n,  \label{82}
\end{equation}
with respect to $(n+1)$ compatible $3n\times 3n$ Poisson structures of rank $%
2n$ 
\begin{gather}
\pi _0 =\left( 
\begin{array}{cccc}
\theta _0 & 0 & \cdots & 0 \\ 
0 &  &  &  \\ 
\vdots &  & 0 &  \\ 
0 &  &  & 
\end{array}
\right) ,\,\,\,\pi _1=\left( 
\begin{array}{ccccc}
\theta _1 & \overline{K}_1 & 0 & \cdots & 0 \\ 
-\overline{K}_1^T &  &  &  &  \\ 
0 &  &  &  &  \\ 
\vdots &  &  & 0 &  \\ 
0 &  &  &  & 
\end{array}
\right) ,  \nonumber \\
\pi _2 =\left( 
\begin{array}{cccccc}
\theta _2 & \overline{K}_2 & \overline{K}_1 & 0 & \cdots & 0 \\ 
-\overline{K}_2^T &  &  &  &  &  \\ 
-\overline{K}_1^T &  &  &  &  &  \\ 
0 &  &  & 0 &  &  \\ 
\vdots &  &  &  &  &  \\ 
0 &  &  &  &  & 
\end{array}
\right) ,\,\,.....,  \label{83} \\
\pi _n =\left( 
\begin{array}{ccccc}
\theta _n & \overline{K}_n & \overline{K}_{n-1} & \cdots & \overline{K}_1 \\ 
-\overline{K}_n^T &  &  &  &  \\ 
-\overline{K}_{n-1}^T &  &  &  &  \\ 
\vdots &  &  & 0 &  \\ 
-\overline{K}_1^T &  &  &  & 
\end{array}
\right) ,  \nonumber
\end{gather}
where 
\begin{equation}
\theta _m=N^m\circ \theta _0=\left( 
\begin{array}{cc}
0 & \Lambda ^m \\ 
-\Lambda ^m & 0
\end{array}
\right) ,\,\,\,\,\Lambda _m=diag(\lambda _1^m,...,\lambda _n^m)  \label{84}
\end{equation}
$\overline{K}_m=(h_{m,\mu} ,-h_{m,\lambda} )^T$ and each Poisson structure $%
\pi _m$ has $n$ Casimir functions: \linebreak[4] $c_{m+1}, c_{m+2}, \ldots ,
c_n, h_n, \ldots , h_{n-m+1}.$ Moreover all functions $h_r(\lambda ,\mu ,c),$
$r=1,...,n$ are in involution with respect to an arbitrary Poisson tensor $%
\pi _k,\,\,k=1,...,n.$

\medskip \ Now, we integrate equations of motion from the hierarchy (\ref{82}%
) solving the Hamilton-Jacobi equation for Hamiltonians (\ref{80}) 
\begin{equation}
h_r(\lambda ,\frac{\partial W}{\partial \lambda },c)=\sum_{k=1}^n\frac{\rho
_{r-1}^k(\lambda )f_k(\lambda _k,\partial W/\partial \lambda _k)}{\Delta _k}%
+\sum_{i=1}^nc_i\beta _{i,r}(\lambda )=a_r.  \label{85}
\end{equation}
First we demonstrate the separability of this equation. Taking the
generating function $W(\lambda ,a)$ in the form $W(\lambda
,a)=\sum_{i=1}^nW_i(\lambda _i,a)$ and the following representation of $%
\beta _{k,r}$%
\begin{equation}
\beta _{k,r}(\lambda )=\sum_{i=1}^n\frac{\partial \rho _r}{\partial \lambda
_i}\frac{\lambda _i^{n+k-1}}{\Delta _i}=-\sum_{i=1}^n\rho _{r-1}^i\frac{%
\lambda _i^{n+k-1}}{\Delta _i},  \label{86}
\end{equation}
eq.(\ref{85}) turns into the form 
\begin{equation}
\sum_{k=1}^n\frac{\rho _{r-1}^k(\lambda )[f(\lambda _k,\partial W_k/\partial
\lambda _k)-\sum_{i=1}^nc_i\lambda _k^{n-1+i}]}{\Delta _k}=a_r.  \label{87}
\end{equation}
Applying relation (\ref{21a}) we get the solution of eq.(\ref{87}) in the
form 
\begin{equation}
f(\lambda _k,\partial W_k/\partial \lambda _k)=g(\lambda _k),\,\,\,k=1,...,n,
\label{88}
\end{equation}
where 
\begin{equation}
g(\xi )=c_n\xi ^{2n-1}+...+c_1\xi ^n+a_1\xi ^{n-1}+...+a_{n-1}\xi +a_n.
\label{89}
\end{equation}
Hence, $W(\lambda ,a)$ can be obtained by solving $n$ decoupled first-order
ODEs (\ref{88}). For example, if 
\begin{equation}
f(\lambda _i,\mu _i)=\varphi (\lambda _i)f(\mu _i)+\psi (\lambda _i),
\label{90}
\end{equation}
then we obtain 
\begin{equation}
W(\lambda ,a)=\sum_{k=1}^n\int^{\lambda _k}f^{-1}\left( \frac{g(\xi )-\psi
(\xi )}{\varphi (\xi )}\right) \,d\xi .  \label{91}
\end{equation}
In new canonical variables $a_i,\,b_i=\frac{\partial W}{\partial a_i},$ the
Hamiltonians $h_r(\lambda ,\mu ,c)$ become $h_r=a_r$ with 
\begin{equation}
b_i=\frac{\partial W}{\partial a_i}=\sum_{k=1}^n\int^{\lambda _k}\left(
f^{-1}\right) ^{\prime }\frac{\xi ^{n-i}}{\varphi (\xi )}\,d\xi ,  \label{92}
\end{equation}
where $\left( f^{-1}\right) ^{\prime }$ means the derivative of\thinspace $%
f^{-1}$. As in the new coordinates each $h_r$ generates a trivial flow 
\begin{equation}
(a_j)_{t_r}=-\frac{\partial h_r}{\partial b_j}=0,\,\,\,(b_j)_{t_r}=\frac{%
\partial h_r}{\partial a_j}=\delta _{j,r},\,\,\,(c_j)_{t_r}=0,  \label{93}
\end{equation}
hence 
\begin{equation}
b_i=t_i+const.  \label{94}
\end{equation}
Combining (\ref{92}) with (\ref{94}) we arrive at implicit solutions for the
trajectories $\lambda _i(t_r),$ with respect to the evolution parameter $t_r$
in the form 
\begin{equation}
\sum_{k=1}^n\int^{\lambda _k}\left( f^{-1}\right) ^{\prime }\frac{\xi ^{n-i}%
}{\varphi (\xi )}\,d\xi =\delta _{i,r}\,t_r+const,\,\,\,\,i=1,...,n.
\label{95}
\end{equation}

\subsection{Multi-Hamiltonian chains in arbitrary canonical coordinates}

Let us introduce arbitrary canonical coordinates $(q,p,c)$ related to the
Darboux-Nijenhuis coordinates $(\lambda ,\mu ,c)$ trough some canonical
transformation 
\begin{equation}
q_k=\zeta _k(\lambda ,\mu ),\,\,\,\,\,\,\,\,p_k=\eta _k(\lambda ,\mu
),\,\,\,\,\,\,\,\,\,\,\,\,\,k=1,...,n.  \label{96}
\end{equation}
Applying the inverse of this transformation to Hamiltonian functions (\ref
{80}) and Poisson matrices (\ref{83}) one finds that 
\begin{equation}
h_r(q,p,c)=h_r(q,p)+\sum_{i=1}^nc_ib_{i,r}(q)  \label{97}
\end{equation}
and the nondegenerate part $\theta _m$ of rank $2n$ of each $\pi _m$(also
implectic) takes now the form 
\begin{equation}
\theta _m=N^m\circ \theta _0=\left( 
\begin{array}{cc}
D_m(q,p) & A_m(q,p) \\ 
-A_m^T(q,p) & B_m(q,p)
\end{array}
\right) ,\,\,\,\,\,\,\,\,\,\,m=1,...,n.  \label{99}
\end{equation}

Conversely, if we have a multi-Hamiltonian chain in $(q,p,c)$ coordinates
and the Nijenhuis tensor 
\begin{equation}
N(q,p)=\left( 
\begin{array}{cc}
D_1(q,p) & A_1(q,p) \\ 
-A_1^T(q,p) & B_1(q,p)
\end{array}
\right) \cdot \left( 
\begin{array}{cc}
0 & I \\ 
-I & 0
\end{array}
\right) ^{-1}=\left( 
\begin{array}{cc}
A_1(q,p) & -D_1(q,p) \\ 
B_1(q,p) & A_1^T(q,p)
\end{array}
\right)  \label{100}
\end{equation}
is nondegenerate and has $n$ distinct eigenvalues $\lambda _i\,$ each of
multiplicity $2$, then the canonical transformation (\ref{96}) transforms a
given chain to the one considered in the previous subsection.

The admissible reductions of the number of Casimir variables are the
following. For arbitrary $1\leq m<n,$ let $c_i\neq 0,$ $1\leq i\leq m$ and $%
c_i=0$ for $m<i\leq n.$ The first $m$ Poisson structures $\pi _i,\,i\leq m$
survive the projection $(\lambda ,\mu ,c_1,...,c_n)\in M\rightarrow 
\overline{M}\ni (\lambda ,\mu ,c_1,...,c_m)$ and we have still $\bigl( 
\begin{smallmatrix}
m+1 \\ 
2
\end{smallmatrix}
\bigr) $ bi-Hamiltonian chains (\ref{82}).\emph{\ }In the limit $%
c_1=...=c_n=0,$ the systems considered lose the bi-Hamiltonian property,
turning into the quasi-bi-Hamiltonian systems on a symplectic manifold $M\ni
(q,p),\,$ being still separable and integrable by quadratures. Moreover,
because of the property (\ref{86}), each of the multi-Hamiltonian systems
considered on a Poisson manifold $M\ni (q,p,c)\,$ has a quasi-bi-Hamiltonian
representation on a symplectic $\,$leaf $S$ of $\pi _0$ ($\dim S=2n$) fixing
the values of all $c_i$.

\subsection{Examples}

The theory extended in this section will be illustrated by several
representative examples of already known as well as new multi-Hamiltonian
systems.

\begin{example}
\emph{Stationary }$t_2-$\emph{flow of dispersive water waves.}\newline
The Hamiltonian functions and Poisson structures in Ostrogradsky variables
are as follows \cite{8}: 
\begin{gather*}
h_1(q,p,c)=-4p_1p_2+5q_2p_1^2-\frac 58q_1q_2^3-\frac 34q_1^2q_2-\frac
7{64}q_2^5+\frac 12q_2c_1+(\frac 12q_1+\frac 18q_2^2)c_2, \\
h_2(q,p,c)=q_1p_1^2+4q_2p_1p_2-\frac 54q_2^2p_1^2-2p_2^2+\frac
5{64}q_1q_2^4-\frac 3{16}q_1^2q_2^2-\frac 14q_1^3+\frac{45}{6\cdot 128}q_2^6
\\
{}\,\,\,\,\,\,\,\,\,\,\,\,\,\,\,\,\,\,\,\,\,\,\,\,\,\,\,\,\,\,\,\,\,\,\,\,\,%
\,\,\,\,\,\,\,\,\,\,\,\,\,\,\,\,\,\,\,\,\,\,\,\,\,\,\,+(\frac 12q_1+\frac
38q_2^2)c_1-(\frac 14q_1q_2-\frac 3{16}q_2^3)c_2,
\end{gather*}
\begin{gather*}
\pi _0=\left( 
\begin{array}{cccccc}
0 & 0 & 1 & 0 & 0 & 0 \\ 
0 & 0 & 0 & 1 & 0 & 0 \\ 
-1 & 0 & 0 & 0 & 0 & 0 \\ 
0 & -1 & 0 & 0 & 0 & 0 \\ 
0 & 0 & 0 & 0 & 0 & 0 \\ 
0 & 0 & 0 & 0 & 0 & 0
\end{array}
\right) , \\
\pi _1=\left( 
\begin{array}{cccccc}
0 & 0 & -\frac 32q_2 & -\frac 12q_1-\frac{15}8q_2^2 & h_{1,p_1} & 0 \\ 
0 & 0 & 1 & q_2 & h_{1,p_2} & 0 \\ 
\frac 32q_2 & -1 & 0 & -p_1 & -h_{1,q_1} & 0 \\ 
\frac 12q_1+\frac{15}8q_2^2 & -q_2 & p_1 & 0 & -h_{1,q_2} & 0 \\ 
-h_{1,p_1} & -h_{1,p_2} & h_{1,q_1} & h_{1,q_2} & 0 & 0 \\ 
0 & 0 & 0 & 0 & 0 & 0
\end{array}
\right) , \\
\pi _2=\left( 
\begin{array}{cccccc}
0 & 0 & \frac 38q_2^2-\frac 12q_1 & -\frac 14q_1q_2-\frac{15}{16}q_2^3 & 
h_{2,p_1} & h_{1,p_1} \\[2pt] 
0 & 0 & -\frac 12q_2 & -\frac 12q_1-\frac 78q_2^2 & h_{2,p_2} & h_{1,p_2} \\%
[2pt] 
-\frac 38q_2^2+\frac 12q_1 & \frac 12q_2 & 0 & \frac 12q_2p_1 & -h_{2,q_1} & 
-h_{1,q_1} \\[2pt] 
\frac 14q_1q_2+\frac{15}{16}q_2^3 & \frac 12q_1+\frac 78q_2^2 & -\frac
12q_2p_1 & 0 & -h_{2,q_2} & -h_{1,q_2} \\[2pt] 
-h_{2,p_1} & -h_{2,p_2} & h_{2,q_1} & h_{2,q_2} & 0 & 0 \\[2pt] 
-h_{1,p_1} & -h_{1,p_2} & h_{1,q_1} & h_{1,q_2} & 0 & 0
\end{array}
\right) .
\end{gather*}
Hence, we have three bi-Hamiltonian chains (\ref{82}) 
\begin{equation}
\begin{split}
& 
\begin{array}{lll}
\pi _0\circ dc_1= & 0 &  \\ 
\pi _0\circ dh_1= & K_1 & =\pi _1\circ dc_1 \\ 
\pi _0\circ dh_2= & K_2 & =\pi _1\circ dh_1 \\ 
& 0 & =\pi _1\circ dh_2,
\end{array}
\,\,\,\,\,\,\,\,\,\,\,\, 
\begin{array}{lll}
\pi _0\circ dc_1= & 0 &  \\ 
\pi _0\circ dh_1= & K_1 & =\pi _2\circ dc_2 \\ 
\pi _0\circ dh_2= & K_2 & =\pi _2\circ dc_1 \\ 
& 0 & =\pi _2\circ dh_1,
\end{array}
\, \\[1em]
& 
\begin{array}{lll}
\pi _1\circ dc_2= & 0 &  \\ 
\pi _1\circ dc_1= & K_1 & =\pi _2\circ dc_2 \\ 
\pi _1\circ dh_1= & K_2 & =\pi _2\circ dc_1 \\ 
& 0 & =\pi _2\circ dh_1.
\end{array}
\end{split}
\label{100a}
\end{equation}
The canonical transformation to the Darboux-Nijenhuis coordinates reads 
\begin{gather*}
q_1=-(3\lambda _1^2+3\lambda _2^2+4\lambda _1\lambda _2), \\
q_2=-2(\lambda _1+\lambda _2), \\
p_1=\frac 12\frac{\mu _2-\mu _1}{\lambda _1-\lambda _2}, \\
p_2=-\frac 12\frac{\lambda _1(3\mu _2-2\mu _1)-\lambda _2(3\mu _1-2\mu _2)}{%
\lambda _1-\lambda _2},
\end{gather*}
where now $h_r,\,r=1,2$ take the form (\ref{80}) with 
\[
f_i(\lambda _i,\mu _i)=2\lambda _i^6-\frac 12\mu _i^2, 
\]
$\beta _{1,1}=\rho _1,\beta _{1,2}=\rho _2$ (\ref{19a}) and 
\begin{equation}
\begin{split}
& \beta _{2,1}=-\lambda _1^2-\lambda _2^2-\lambda _1\lambda _2, \\
& \beta _{2,2}=\lambda _1\lambda _2(\lambda _1+\lambda _2).
\end{split}
\label{102}
\end{equation}
\end{example}

\begin{example}
\emph{Two-Casimir extension of the Henon-Heiles system.} \newline
In Example 1 one-Casimir extension of the Henon-Heiles system was considered
in the form 
\[
(q_1)_{tt}=-3q_1^2-\frac 12q_2^2+c,\,\,\,\,\,\,\,\,(q_2)_{tt}=-q_1q_2, 
\]
with two constants of motion (\ref{52}) and the related transformation to
Darboux-Nijenhuis coordinates (\ref{58a}). For a two-Casimir extension we
get immediately 
\begin{equation}
\begin{split}
& \beta _{2,1}=-[(\lambda _1+\lambda _2)^2-\lambda _1\lambda
_2]=-(q_1^2+\frac 14q_2^2), \\
& \beta _{2,2}=\lambda _1\lambda _2(\lambda _1+\lambda _2)=-\frac 14q_1q_2^2.
\end{split}
\label{103}
\end{equation}
Hence 
\begin{equation}
\begin{split}
& h_1=H=\frac 12p_1^2+\frac 12p_2^2+q_1^3+\frac
12q_1q_2^2-c_1q_1-(q_1^2+\frac 14q_2^2)c_2, \\
& h_2=\frac 12q_2p_1p_2-\frac 12q_1p_2^2+\frac 1{16}q_2^4+\frac
14q_1^2q_2^2-\frac 14q_2^2c_1-\frac 14q_1qc_2,
\end{split}
\label{104}
\end{equation}
where the Newton's equations related to the energy $H$ are 
\begin{equation}
(q_1)_{tt}=-3q_1^2-\frac
12q_2^2+c_1+2q_1c_2,\,\,\,\,\,\,\,\,(q_2)_{tt}=-q_1q_2+\frac 12q_2c_2.
\label{105}
\end{equation}
This is tri-Hamiltonian system with the following Poisson structures 
\begin{gather*}
\pi _0=\left( 
\begin{array}{cccccc}
0 & 0 & 1 & 0 & 0 & 0 \\ 
0 & 0 & 0 & 1 & 0 & 0 \\ 
-1 & 0 & 0 & 0 & 0 & 0 \\ 
0 & -1 & 0 & 0 & 0 & 0 \\ 
0 & 0 & 0 & 0 & 0 & 0 \\ 
0 & 0 & 0 & 0 & 0 & 0
\end{array}
\right) , \\
\pi _1=\left( 
\begin{array}{cccccc}
0 & 0 & q_1 & \frac 12q_2 & h_{1,p_1} & 0 \\ 
0 & 0 & \frac 12q_2 & 0 & h_{1,p_2} & 0 \\ 
-q_1 & -\frac 12q_2 & 0 & \frac 12p_2 & -h_{1,q_1} & 0 \\ 
-\frac 12q_2 & 0 & -\frac 12p_2 & 0 & -h_{1,q_2} & 0 \\ 
-h_{1,p_1} & -h_{1,p_2} & h_{1,q_1} & h_{1,q_2} & 0 & 0 \\ 
0 & 0 & 0 & 0 & 0 & 0
\end{array}
\right) , \\
\pi _2=\left( 
\begin{array}{cccccc}
0 & 0 & q_1^2+\frac 14q_2^2 & \frac 12q_1q_2 & h_{2,p_1} & h_{1,p_1} \\[2pt] 
0 & 0 & \frac 12q_1q_2 & \frac 14q_2^2 & h_{2,p_2} & h_{1,p_2} \\[2pt] 
-q_1^2-\frac 14q_2^2 & -\frac 12q_1q_2 & 0 & \frac 12q_1p_2 & -h_{2,q_1} & 
-h_{1,q_1} \\[2pt] 
-\frac 12q_1q_2 & -\frac 14q_2^2 & -\frac 12q_1p_2 & 0 & -h_{2,q_2} & 
-h_{1,q_2} \\[2pt] 
-h_{2,p_1} & -h_{2,p_2} & h_{2,q_1} & h_{2,q_2} & 0 & 0 \\[2pt] 
-h_{1,p_1} & -h_{1,p_2} & h_{1,q_1} & h_{1,q_2} & 0 & 0
\end{array}
\right) .
\end{gather*}
The first two of them come from one-Casimir extension and the last one was
constructed according to formula (\ref{99}). Notice that again we have three
bi-Hamiltonian chains (\ref{100a}).

Also in Example 1 the inverse one-Casimir extension of the Henon-Heiles
system was considered in the form 
\[
(q_1)_{tt}=-3q_1^2-\frac 12q_2^2,\,\,\,\,\,\,\,\,(q_2)_{tt}=-q_1q_2-\frac
8{q_2^3}c, 
\]
with two constants of motion (\ref{55}) and the respective transformation to
the DN coordinates (\ref{59a}). The two-Casimir extension we get by adding
new terms 
\begin{gather*}
\beta _{2,1}=-[(\lambda _1+\lambda _2)^2-\lambda _1\lambda _2]=\frac
4{q_2^2}-\frac{16q_1^2}{q_2^4}, \\
\beta _{2,2}=\lambda _1\lambda _2(\lambda _1+\lambda _2)=\frac{16q_1}{q_2^4},
\end{gather*}
to the constants of motion (\ref{55}) 
\begin{gather*}
h_1^{\prime }(q,p,c_1,c_2)=h_1^{\prime }(q,p,c=c_1)+\frac{16q_1}{q_2^4}\,c_2,
\\
h_2^{\prime }(q,p,c_1,c_2)=h_2^{\prime }(q,p,c=c_1)+\left( \frac 4{q_2^2}-%
\frac{16q_1^2}{q_2^4}\right) c_2,
\end{gather*}
where the Newton equations related to the energy $h_1^{\prime }$ are 
\[
(q_1)_{tt}=-3q_1^2-\frac 12q_2^2-\frac{16}{q_2^4}c_2,\,\,\,\,\,\,\,%
\,(q_2)_{tt}=-q_1q_2-\frac 8{q_2^3}c_1+\frac{64q_1}{q_2^5}c_2. 
\]
Again this is inverse tri-Hamiltonian system 
\begin{equation}
\begin{split}
& 
\begin{array}{l@{\ }l@{\ }l}
\pi _0\circ dc_1= & 0 &  \\ 
\pi _0\circ dh_2^{\prime }= & K_2^{\prime } & =\pi _{-1}\circ dc_1 \\ 
\pi _0\circ dh_1^{\prime }= & K_1^{\prime } & =\pi _{-1}\circ dh_2^{\prime }
\\ 
& 0 & =\pi _{-1}\circ dh_1^{\prime },
\end{array}
\,\,\,\,\,\,\,\,\,\,\,\, 
\begin{array}{l@{\ }l@{\ }l}
\pi _0\circ dc_1= & 0 &  \\ 
\pi _0\circ dh_2^{\prime }= & K_2^{\prime } & =\pi _{-2}\circ dc_2 \\ 
\pi _0\circ dh_1^{\prime }= & K_1^{\prime } & =\pi _{-2}\circ dc_1 \\ 
& 0 & =\pi _{-2}\circ dh_2^{\prime },
\end{array}
\, \\[1em]
& 
\begin{array}{l@{\ }l@{\ }l}
\pi _{-1}\circ dc_2= & 0 &  \\ 
\pi _{-1}\circ dc_1= & K_2^{\prime } & =\pi _{-2}\circ dc_2 \\ 
\pi _{-1}\circ dh_2^{\prime }= & K_1^{\prime } & =\pi _{-2}\circ dc_1 \\ 
& 0 & =\pi _{-2}\circ dh_2^{\prime },
\end{array}
\end{split}
\label{106}
\end{equation}
where the Poisson structure $\pi _{-1}$is given by (\ref{56}), with
additional last row and column with zeros, while the new third Poisson
structure $\pi _{-2}$ reads 
\[
\pi _{-2}=\left( 
\begin{array}{cccccc}
0 & 0 & 4/q_2^2 & -8q_{/}q_2^3 & h_{1,p_1}^{\prime } & h_{2,p_1}^{\prime } \\%
[2pt] 
0 & 0 & -8q_{/}q_2^3 & (16q_1^2+4q_2^2)/q_2^4 & h_{1,p_2}^{\prime } & 
h_{2,p_2}^{\prime } \\[2pt] 
-4/q_2^2 & 8q_1/q_2^3 & 0 & -8q_1p_2/q_2^4 & -h_{1,q_1}^{\prime } & 
-h_{2,q_1}^{\prime } \\[2pt] 
8q_1/q_2^3 & -(16q_1^2+4q_2^2)/q_2^4 & 8q_1p_2/q_2^4 & 0 & 
-h_{1,q_2}^{\prime } & -h_{2,q_2}^{\prime } \\[2pt] 
-h_{1,p_1}^{\prime } & -h_{1,p_2}^{\prime } & h_{1,q_1}^{\prime } & 
h_{1,q_2}^{\prime } & 0 & 0 \\[2pt] 
-h_{2,p_1}^{\prime } & -h_{2,p_2}^{\prime } & h_{2,q_1}^{\prime } & 
h_{2,q_2}^{\prime } & 0 & 0
\end{array}
\right) . 
\]
\end{example}

Till now all examples presented were bi(multi)-Hamiltonian St\"{a}ckel
systems \cite{2}, i.e. the systems with all Hamiltonian functions quadratic
in momenta in DN coordinates: $f(\lambda _i,\mu _i)=\varphi (\lambda _i)\mu
_i^2+\psi (\lambda _i).$ Here we present the first example of
non-St\"{a}ckel system.

\begin{example}
\emph{m-Casimir extension of the relativistic $n$-body problem.} \newline
Consider the Hamiltonian dynamical system with the Hamiltonian given by 
\begin{equation}
H=\sum_{i=1}^n\frac{\varphi _i(\lambda _i)}{\Delta _i}e^{a\mu _i},
\label{107}
\end{equation}
where $\varphi _i$ are arbitrary smooth functions and $a$ is an arbitrary
constant. The corresponding dynamical system takes the form 
\begin{equation}
(\lambda _i)_{tt}=2\sum_{k\neq i}\frac{(\lambda _i)_t(\lambda _k)_t}{\lambda
_i-\lambda _k},\,\,\,\,\,\,\,\,\,i=1,...,n,  \label{108}
\end{equation}
which depends explicitly on velocities. The derivation of formula (\ref{108}%
) is given in ref. \cite{1}. Notice that equations (\ref{108}) do not depend
on $\varphi _i$ functions, hence the dynamics is not influenced by $\varphi
_i$ terms.

Dynamics (\ref{108}) is a special case of the integrable relativistic $n$%
-body problems introduced by Ruijsenaars and Schneider \cite{9}. Now,
comparing (\ref{107}) with (\ref{80}) one immediately concludes that $%
(\lambda ,\mu )$ is a Darboux-Nijenhuis chart for the Hamiltonian $H$, and
as a consequence, system (\ref{108}) is quasi-bi-Hamiltonian and separable,
with the solution given by the implicit formulae 
\begin{equation}
\frac 1a\sum_{k=1}^n\int^{\lambda _k}\frac{\xi ^{n-i}}{g(\xi )}\,d\xi
=\delta _{i,1}\,t+const,\,\,\,\,i=1,...,n,  \label{109}
\end{equation}
where $g(\xi )=a_n+a_{n-1}\xi +...+a_1\xi ^{n-1}.$ This fact was noticed for
the first time by Morosi and Tondo \cite{mor2}. Notice, that trajectories $%
\lambda _i(t)$ do not depend on the $\varphi _i(\lambda _i)$ factors, as
expected, so without a loss of generality one can put $\varphi _i(\lambda
_i)=1.$

The system (\ref{108}) can be naturally extended to an $m$-Casimir one with
the Hamiltonian 
\begin{equation}
H=\sum_{i=1}^n\frac{\varphi _i(\lambda _i)}{\Delta _i}e^{a\mu
_i}+\sum_{j=1}^mc_j\beta _{j,1}(\lambda ),\,\,\,\,\,\,\,\,\,1\leq m\leq n
\label{110}
\end{equation}
and the related Newton equations of motion 
\begin{equation}
(\lambda _i)_{tt}=2\sum_{k\neq i}\frac{(\lambda _i)_t(\lambda _k)_t}{\lambda
_i-\lambda _k}-a\sum_{j=1}^mc_j\frac{\partial \beta _{j,1}}{\partial \lambda
_i}\,(\lambda _i)_t.  \label{111}
\end{equation}
The dynamical system (\ref{111}) has $n$ constants of motion 
\[
h_r(\lambda ,\mu ,c)=-\sum_{i=1}^n\frac{\partial \rho _r(\lambda )}{\partial
\lambda _i}\frac{e^{a\mu _i}}{\Delta _i}+\sum_{j=1}^mc_j\beta _{j,r}(\lambda
),\,\,\,\,r=1,...,n, 
\]
$(m+1)$ Poisson structures (\ref{83}) and the solution given by implicit
formula (\ref{109}), where now 
\[
g(\xi )=a_n+a_{n-1}\xi +...+a_1\xi ^{n-1}+c_1\xi ^n+...+c_m\xi ^{n+m-1}. 
\]
The one-Casimir extension, which is bi-Hamiltonian, has the following Newton
equations of motion 
\[
(\lambda _i)_{tt}=2\sum_{k\neq i}\frac{(\lambda _i)_t(\lambda _k)_t}{\lambda
_i-\lambda _k}+ac(\lambda _i)_t,\,\,\,\,\,\,\,\,\,i=1,...,n. 
\]
The two-Casimir extension, which is tri-Hamiltonian, has the Newton
equations in the form 
\[
(\lambda _i)_{tt}=2\sum_{k\neq i}\frac{(\lambda _i)_t(\lambda _k)_t}{\lambda
_i-\lambda _k}+ac_1(\lambda _i)_t+ac_2\left( \lambda _i+\sum_{k=1}^n\lambda
_k\right) (\lambda _i)_t,\,\,\,\,\,\,\,\,\,i=1,...,n. 
\]
\end{example}

\begin{example}
\emph{$(m+1)$-Hamiltonian formulation for elliptic separable potentials.} 
\newline
In Example 3 we presented bi-Hamiltonian systems separable in generalized
elliptic coordinates. In the following example we extend the result onto
appropriate multi-Hamiltonian chains.

According to the theory presented in this section, let us generalize the
Hamiltonian system (\ref{67}) to the form 
\begin{equation}
H(q,p,c)=\frac
12(p,p)+V(q)+\sum_{k=1}^mc_kb_{k,1}(q),\,\,\,\,\,\,\,\,\,\,m=1,...,n
\label{112}
\end{equation}
being multi-Hamiltonian and separable. The few first $b_{k,1}(q)$ functions
are as follows 
\begin{equation}
\begin{split}
& b_{1,1}(q)=\frac 12(q,q), \\
& b_{2,1}(q)=\frac 12(q,Aq)-\frac 14(q,q)^2, \\
& b_{3,1}(q)=\frac 12(q,q)(q,Aq)-\frac 12(q,A^2q)-\frac 18(q,q)^3,...\,\,.
\end{split}
\label{113}
\end{equation}

For example, the three Poisson structures of the system (\ref{112}) with two
Casimirs read 
\begin{gather*}
\pi _0=\left( 
\begin{array}{cccc}
0 & I & 0 & 0 \\ 
-I & 0 & 0 & 0 \\ 
0 & 0 & 0 & 0 \\ 
0 & 0 & 0 & 0
\end{array}
\right) , \\
\pi _1=\left( 
\begin{array}{cccc}
0 & A-\frac 12q\otimes q & h_{1,p} & 0 \\[2pt] 
-A+\frac 12q\otimes q & \frac 12p\otimes q-\frac 12q\otimes p & -h_{1,q} & 0
\\[2pt] 
-(h_{1,p})^T & (h_{1,q})^T & 0 & 0 \\[2pt] 
0 & 0 & 0 & 0
\end{array}
\right) , \\
\pi _2=\left( 
\begin{array}{cccc}
0 & (A-\frac 12q\otimes q)^2 & h_{2,p} & h_{1,p} \\[1ex] 
-(A-\frac 12q\otimes q)^2 & 
\begin{array}{c}
(A-\frac 12q\otimes q)(\frac 12p\otimes q-\frac 12q\otimes p) \\ 
+(\frac 12p\otimes q-\frac 12q\otimes p)(A-\frac 12q\otimes q)
\end{array}
& -h_{2,q} & -h_{1,q} \\[2.5ex] 
-(h_{2,p})^T & (h_{2,q})^T & 0 & 0 \\[2pt] 
-(h_{1,p})^T & (h_{1,q})^T & 0 & 0
\end{array}
\right) .
\end{gather*}
The functions $h_r(q,p,c),$ forming three admissible bi-Hamiltonian chains,
are given by formulas (\ref{70}), where now 
\[
\overline{h}_r=\frac 12\sum_{i=1}^n\alpha
_i^{r-1}K_i+c_1b_{1,r}(q)+c_2b_{2,r}(q). 
\]
\end{example}

\section{Multi-Casimir split chains}

In the following section we extend the results from previous sections onto
the so-called split bi(multi)-Hamiltonian chains \cite{6}. Potentially, the
variety of such systems is much richer than the one of unsplit chains, but
still less recognized. The reason is that systems from this family are
generally non-physical in the sense that are either non-St\"{a}ckel, or even
if are of St\"{a}ckel type, the underlying St\"{a}ckel space is never flat
(at most conformally flat).

Let $M$ be a $(2n+k)$ dimensional manifold endowed with a linear Poisson
pencil $\pi _\lambda =\pi _1-\lambda \pi _0.$ We suppose that it admits $k$
polynomial Casimir functions 
\begin{equation}
h_\lambda ^{(\alpha )}=\sum_{j=0}^{n_\alpha }h_j^{(\alpha )}\lambda
^{n_\alpha -j},\,\,\,\,\,\,\,\,\,\,\alpha =1,...,k  \label{113a}
\end{equation}
with $n=n_1+...+n_k.$ From beeing Casimir of a pencil it follows that the
set $\{h_0^{(\alpha )}\}_{\alpha =1}^k$ is a set of Casimirs of $\pi _0,$
while $\{h_{n_\alpha }^{(\alpha )}\}_{\alpha =1}^k$ is a set of Casimirs of $%
\pi _1,$ respectively. In canonical coordinates $(q,p,c)$ $h_0^{(\alpha
)}=c_\alpha ,\,\alpha =1,...,k,$ 
\begin{equation}
\pi _0=\left( 
\begin{array}{cccc}
\theta _0 & 0 & ... & 0 \\ 
0 &  &  &  \\ 
\vdots &  & 0 &  \\ 
0 &  &  & 
\end{array}
\right) ,\,\,\,\pi _1=\left( 
\begin{array}{cccc}
\theta _1 & K_1^{(1)} & \cdots & K_1^{(k)} \\ 
-\left( K_1^{(1)}\right) ^T &  &  &  \\ 
\vdots &  & 0 &  \\ 
-\left( K_1^{(k)}\right) ^T &  &  & 
\end{array}
\right) ,  \label{113b}
\end{equation}
where $K_1^{(i)}=\theta _0dh_1^{(i)}$ and $\theta _0,\theta _1$ are given by
(\ref{31}). Now, we are looking for a separation curve in the form 
\begin{equation}
f(\lambda ,\mu )=\sum_\alpha \vartheta _\alpha h_\lambda ^{(\alpha
)}=h_\lambda ,  \label{113c}
\end{equation}
where $\vartheta _\alpha $ are admissible functions of $\lambda $ and $\mu .$
Further on we concentrate on particular two-Casimir cases, but it will be
enough to make some insight into the theory.

Let us start from a separation curve for the unsplit two-Casimir case 
\begin{equation}
f(\lambda ,\mu )=h_\lambda ,\,\,\,\,h_\lambda =c_2\lambda ^{n+1}+c_1\lambda
^n+h_1\lambda ^{n-1}+...+h_n.  \label{114}
\end{equation}
An arbitrary shift of $c_1$ Casimir variable along the polynomial $h_\lambda 
$ leads to a separation curve 
\begin{equation}
f(\lambda ,\mu )=c_2\lambda ^{n+1}+h_1\lambda ^n+...+h_i\lambda
^{n-i+1}+c_1\lambda ^{n-i}+h_{i+1}\lambda ^{n-i-1}+...+h_n  \label{115}
\end{equation}
of some split bi-Hamiltonian chain. This is the case (\ref{113c}) with $%
k=2,\,n_1=n-i,\,n_2=i,\,\vartheta _1=1,\,\vartheta _2=\lambda
^{n+1-i},\,h_l^{(2)}=h_l,l=1,...,i$ and $h_j^{(1)}=h_{i+j},$\thinspace $%
j=1,...,n-i.$ We illustrate the situation for $i=1.$ The following results
were obtained. For a separation curve in the form 
\begin{equation}
f(\lambda ,\mu )=c_2\lambda ^{n+1}+h_1\lambda ^n+c_1\lambda
^{n-1}+h_2\lambda ^{n-2}+...+h_n\equiv h_\lambda   \label{116}
\end{equation}
Hamiltonian functions $h_r,r=1,...,n$ read 
\begin{equation}
h_r(\lambda ,\mu ,c)=h_r(\lambda ,\mu )+\gamma _{1,r}(\lambda )c_1+\gamma
_{2,r}(\lambda )c_2,  \label{117}
\end{equation}
where 

\begin{equation}
\begin{split}
& h_r(\lambda ,\mu )=\sum_{i=1}^n\alpha _{r-1}^i(\lambda )\frac{f(\lambda
_i,\mu _i)}{\Omega _i(\lambda )}, \\
& \Omega _i(\lambda )=\left( \sum_{k=1}^n\lambda _k\right) \prod_{k\neq
i}(\lambda _i-\lambda _k), \\
& \gamma _{1,1}(\lambda )=\frac 1{\rho _1},\,\,\,\gamma _{1,r}(\lambda )=%
\frac{\rho _r}{\rho _1},\,\,\,r=2,...,n, \\
& \gamma _{2,1}(\lambda )=\frac{\rho _1^2-\rho _2}{\rho _1},\,\,\,\gamma
_{2,r}(\lambda )=\frac{\rho _1\rho _{r+1}-\rho _2\rho _r}{\rho _1}%
,\,\,\,r=2,...,n, \\
& \alpha _r^i(\lambda )=\alpha _r(\lambda _i=0),\,\,\,\alpha _r(\lambda
)\equiv \beta _{2,r}(\lambda )=\rho _{r+1}-\rho _1\rho _r.
\end{split}
\label{118}
\end{equation}
Two compatible Poisson structures 
\begin{equation}
\begin{split}
& \pi _0=\left( 
\begin{array}{cccc}
0 & I & 0 & 0 \\ 
-I & 0 & 0 & 0 \\ 
0 & 0 & 0 & 0 \\ 
0 & 0 & 0 & 0
\end{array}
\right) , \\
& \pi _1=\left( 
\begin{array}{cccc}
0 & \Lambda  & h_{2,\mu } & h_{1,\mu } \\ 
-\Lambda  & 0 & -h_{2,\lambda } & -h_{1,\lambda } \\ 
-(h_{2,\mu })^T & (h_{2,\lambda })^T & 0 & 0 \\ 
-(h_{1,\mu })^T & (h_{1,\lambda })^T & 0 & 0
\end{array}
\right) ,
\end{split}
\label{119}
\end{equation}
give rise to a linear Poisson pencil $\pi _\lambda =\pi _1-\lambda \pi _0$,
which acting on its Casimir $h_\lambda $ generates a bi-Hamiltonian chain 
\begin{equation}
\pi _\lambda \circ dh_\lambda \,\,\Longrightarrow \,\,
\begin{array}{l@{\ }l@{\ }l}
\pi _0\circ dc_2= & 0 &  \\ 
\pi _0\circ dh_1= & K_1 & =\pi _1\circ dc_2 \\ 
& 0 & =\pi _1\circ dh_1 \\ 
\pi _0\circ dc_1= & 0 &  \\ 
\pi _0\circ dh_2= & K_2 & =\pi _1\circ dc_1 \\ 
& \vdots  &  \\ 
\pi _0\circ dh_n= & K_n & =\pi _1\circ dh_{n-1} \\ 
& 0 & =\pi _1\circ dh_n
\end{array}
\label{120}
\end{equation}
which splits onto two sub-chains, each starting and terminating with a
Casimir of an appropriate Poisson structure.

\begin{example}
\emph{The case of three degrees of freedom.} \newline
We construct a system comparable with the Newton representation of the $7th$
order KdV \cite{1}, \cite{3}. Thus let us take 
\begin{equation}
f(\mu _i,\lambda _i)=\frac 18\mu _i^2+16\lambda _i^7,\,\,\,i=1,2,3
\label{121}
\end{equation}
and the point transformation generated by relations 
\begin{equation}
\begin{split}
& q_1=\lambda _1+\lambda _2+\lambda _3, \\
& q_2=-\frac 12(\lambda _1^2+\lambda _2^2+\lambda _3^2)+(\lambda _1\lambda
_2+\lambda _1\lambda _3+\lambda _2\lambda _3), \\
& q_3=\frac 12(\lambda _1-\lambda _2-\lambda _3)(\lambda _2-\lambda
_1-\lambda _3)(\lambda _3-\lambda _1-\lambda _2).
\end{split}
\label{121a}
\end{equation}
Then, from (\ref{118}), we arrive at the following Hamiltonians in natural
coordinates 
\begin{align*}
h_1={}& \frac 1{q_1}\left( \frac 12p_2^2+p_1p_3\right)
+10q_1q_3+8q_2^2-10q_1^2q_2+3q_1^4-4\frac{q_2q_3}{q_1}-\frac 1{q_1}c_1 \\
& +\left( \frac 12\frac{q_2}{q_1}-\frac 34q_1\right) c_2, \\
h_2={}& \frac 18\left( \frac{2q_2}{q_1}-3q_1\right) p_2^2+\frac
12q_3p_3^2+\frac 14\left( \frac{2q_2}{q_1}-q_1\right) p_1p_3-\frac
12p_1p_2-\frac 12q_2p_2p_3 \\
& -2q_1q_2q_3-q_1^2q_2^2+\frac 32q_1^4q_2+2q_2^3+\frac
52q_1^3q_3+q_3^2-\frac 12q_1^6-2\frac{q_2^2q_3}{q_1} \\
& -\frac 12\left( \frac 12q_1+\frac{q_2}{q_1}\right) c_1+\frac 12\left(
q_1q_2-\frac 12q_3+\frac 18q_1^3+\frac 12\frac{q_2^2}{q_1}\right) c_2, \\
h_3={}& \frac 18p_1^2+\frac 18\left( q_1^2-q_2-\frac{q_3}{q_1}\right)
p_2^2+\frac 18q_2^2p_3^2+\frac 14q_1p_1p_2-\frac 14\frac{q_3}{q_1}p_1p_3 \\
& -\frac 14\left( 2q_3+q_1q_2\right) +\frac 12q_1^4q_3-q_1^2q_2q_3-\frac
12q_1q_3^2+\frac 12q_1^5q_2-\frac 12q_1^3q_2^2-q_1q_2^3 \\
& +\frac{q_2q_3^2}{q_1}+\frac 14\left( q_2-\frac{q_3}{q_1}\right) c_1+\frac
1{16}(q_1^2q_2-q_1q_3+2q_2^2-2\frac{q_2q_3}{q_1})c_2.
\end{align*}
They form bi-Hamiltonian split chain (\ref{120}), where now 
\[
\pi _1=\left( 
\begin{array}{cccccccc}
0 & 0 & 0 & \frac 12q_1 & -\frac 12 & 0 & h_{2,p_1} & h_{1,p_1} \\[2pt]
0 & 0 & 0 & \frac 12q_2 & 0 & -\frac 12 & h_{2,p_2} & h_{1,p_2} \\[2pt]
0 & 0 & 0 & q_3 & \frac 12q_2 & \frac 12q_1 & h_{2,p_3} & h_{1,p_3} \\[2pt]
-\frac 12q_1 & -\frac 12q_2 & -q_3 & 0 & \frac 12p_2 & \frac 12p_3 & 
-h_{2,q_1} & -h_{1,q_1} \\[2pt]
\frac 12 & 0 & -\frac 12q_2 & -\frac 12p_2 & 0 & 0 & -h_{2,q_2} & -h_{1,q_2}
\\[2pt]
0 & \frac 12 & -\frac 12q_1 & -\frac 12p_3 & 0 & 0 & -h_{2,q_3} & -h_{1,q_3}
\\[2pt]
-h_{2,p_1} & -h_{2,p_2} & -h_{2,p_3} & h_{2,q_1} & h_{2,q_2} & h_{2,q_3} & 0
& 0 \\[2pt]
-h_{1,p_1} & -h_{1,p_2} & -h_{1,p_3} & h_{1,q_1} & h_{1,q_2} & h_{1,q_3} & 0
& 0
\end{array}
\right) ,
\]
and are conformally related to these of first Newton representation of $7th$
order stationary KdV, i.e. in separated coordinates the metrices of
respective St\"{a}ckel spaces are conformally related.
\end{example}

Another admissible form of the separation curve with two Casimirs, leading
to bi-Hamiltonian split chain of non-St\"{a}ckel type, reads 
\[
f(\lambda ,\mu )=\mu \left( c_2\lambda ^{n-i}+h_1\lambda
^{n-i-1}+...+h_{n-i}\right) +c_1\lambda ^i+h_{n-i-1}\lambda ^{i-1}+...+h_n. 
\]
It has the form (\ref{113c}) with $k=2,\,n_1=i,\,n_2=n-i,$\thinspace $%
\vartheta _1=1,\,\vartheta _2=\mu ,$ \thinspace $h_l^{(2)}=h_l,l=1,...,n-i,$
and $h_j^{(1)}=h_{n-i+j},\,j=1,...,i.$ $\,$Here we concentrate on the case $%
i=n-1$ 
\begin{equation}
f(\lambda ,\mu )=\mu (c_2\lambda +h_1)+c_1\lambda ^{n-1}+h_2\lambda
^{n-2}+...+h_n.  \label{123}
\end{equation}
The Poisson pencil and the chain are the same as in the previous split case,
i.e. are of the form (\ref{119}) and (\ref{120}), but $h_r(\lambda ,\mu ,c),$
being the solution of the system 
\[
f(\lambda _i,\mu _i)=\mu _i(c_2\lambda _i+h_1)+c_1\lambda
_i^{n-1}+h_2\lambda _i^{n-2}+...+h_n,\,\,\,i=1,...,n, 
\]
are of more complicated form and we omit here the explicit formulas.

We illustrate the case (\ref{123}) and $n=3$ with the example where the
transformation to DN coordinates will be of non-point nature.

\begin{example}
\emph{The 4th order Boussinesq stationary flow. } \newline
The Hamiltonians and compatible Poisson structures for the $4th$ order
Boussinesq stationary flow, written in generalized Ostrogradsky canonical
coordinates, are the following \cite{for} 
\begin{align*}
h_1={}& -27p_2p_3-27p_2^2-9p_3^2+2q_1^2p_3+3q_1^2p_2+q_2p_1-\frac
29q_1q_3^2-\frac 19q_1q_2^2-\frac 8{81}q_1^4 \\
& +\frac 13q_1c_1+\frac 13q_3c_2, \\
h_2={}& -9p_1p_3+q_1^2p_1+2q_1(q_2-2q_3)p_2+\frac 43q_1q_2p_3-\frac
4{27}q_3^3+\frac 8{81}q_1^3q_3-\frac{16}{81}q_1^3q_2 \\
& +\frac 1{27}q_2^3+\frac 29q_2q_3^2-\frac 4{27}q_2^2q_3+\frac
13(2q_3-q_2)c_1+(3p_2-\frac 19q_1^2)c_2, \\
h_3={}&
-54p_2^3-27p_2p_3^2-81p_2^2p_3+3q_1p_1^2+18q_1^2p_2^2+2q_1^2p_3^2+3q_2p_1p_2+3(q_2-q_3)p_1p_3
\\
& +15q_1^2p_2p_3-q_1^2(q_3+\frac 23q_2)p_1-(\frac{44}{27}q_1^4+\frac
23q_1q_2q_3)p_2+(\frac 23q_1q_3^2-\frac 19q_1q_2^2 \\
& -\frac 49q_1q_2q_3-\frac{16}{27}q_1^4)p_3+\frac{32}{243}q_1^3q_2q_3+\frac{%
32}{729}q_1^6+\frac{16}{243}q_1^3q_3^3+\frac 4{243}q_1^3q_2^2 \\
& -\frac 4{81}q_2^2q_3^2+\frac 1{81}q_2^3q_3+\frac 2{27}q_2q_3^3-\frac
1{27}q_3^4+(\frac 4{27}q_1^3+\frac 19q_3^2-\frac 19q_2q_3-2q_1p_2 \\
& -q_1p_3)c_1+(q_1p_1+q_3p_2-\frac 5{27}q_1^2q_3-\frac 2{27}q_1^2q_2)c_2,
\end{align*}
\begin{gather*}
\pi _0=\left( 
\begin{array}{cccccccc}
0 & 0 & 0 & 1 & 0 & 0 & 0 & 0 \\ 
0 & 0 & 0 & 0 & 1 & 0 & 0 & 0 \\ 
0 & 0 & 0 & 0 & 0 & 1 & 0 & 0 \\ 
-1 & 0 & 0 & 0 & 0 & 0 & 0 & 0 \\ 
0 & -1 & 0 & 0 & 0 & 0 & 0 & 0 \\ 
0 & 0 & -1 & 0 & 0 & 0 & 0 & 0 \\ 
0 & 0 & 0 & 0 & 0 & 0 & 0 & 0 \\ 
0 & 0 & 0 & 0 & 0 & 0 & 0 & 0
\end{array}
\right) , \\
\pi _1=\left( 
\begin{array}{cccccccc}
0 & 0 & 0 & -\frac 13q_3 & \frac 13q_1 & -\frac 23q_1 & h_{2,p_1} & h_{1,p_1}
\\[2pt]
0 & 0 & -3q_1 & A & \frac 13q_2-\frac 13q_3 & -\frac 13q_2 & h_{2,p_2} & 
h_{1,p_2} \\[2pt]
0 & 3q_1 & 0 & \frac 29q_1^2 & 0 & -\frac 13q_3 & h_{2,p_3} & h_{1,p_3} \\%
[2pt]
\frac 13q_3 & -A & -\frac 29q_1^2 & 0 & B & C & -h_{2,q_1} & -h_{1,q_1} \\%
[2pt]
-\frac 13q_1 & -\frac 13q_2+\frac 13q_3 & 0 & -B & 0 & -\frac 8{81}q_1^2 & 
-h_{2,q_2} & -h_{1,q_2} \\[2pt]
\frac 23q_1 & \frac 13q_2 & \frac 13q_3 & -C & \frac 8{81}q_1^2 & 0 & 
-h_{2,q_3} & -h_{1,q_3} \\[2pt]
-h_{2,p_1} & -h_{2,p_2} & -h_{2,p_3} & h_{2,q_1} & h_{2,q_2} & h_{2,q_3} & 0
& 0 \\[2pt]
-h_{1,p_1} & -h_{1,p_2} & -h_{1,p_3} & h_{1,q_1} & h_{1,q_2} & h_{1,q_3} & 0
& 0
\end{array}
\right) 
\end{gather*}
where $A=9p_2+3p_3-q_1^2,\,B=\frac 2{27}q_1q_2+\frac 8{81}q_1q_3-\frac
13p_1,\,C=-\frac 8{81}q_1q_2-\frac 4{27}q_1q_3+\frac 13p_1.$

>From the minimal polynomial (\ref{26a}) of related Nijenhuis tensor $N$ one
finds the first part of transformation to the HF coordinates $(u,v)$ 
\begin{equation}
\begin{split}
& u_1=q_3-\frac 13q_2, \\
& u_2=\frac 13q_3^2+\frac 5{27}q_1^3-3q_1p_2-q_1p_3-\frac 29q_2q_3, \\
& u_3=\frac 1{27}q_3^3+\frac 19q_1^3q_3-q_1q_3p_2-\frac 13q_1q_3p_3+\frac
2{81}q_1^3q_2-\frac 13q_1^2p_1-\frac 1{27}q_2q_3,
\end{split}
\label{124}
\end{equation}
while the system (\ref{37}) with the second chain (\ref{35}) give the second
part of the transformation 
\begin{equation}
\begin{split}
& v_1=-\frac 3{q_1}, \\
& v_2=\frac{q_2-2q_3}{q_1}, \\
& v_3=3p_3+6p_2-\frac 13\frac{q_3^2}{q_1}+\frac 13\frac{q_2q_3}{q_1}-\frac
49q_1^2.
\end{split}
\label{125}
\end{equation}
On the other hand the relation between HF and DN coordinates reads 
\begin{equation}
\begin{split}
& u_1=-\lambda _1-\lambda _2-\lambda _3, \\
& u_2=\lambda _1\lambda _2+\lambda _1\lambda _3+\lambda _2\lambda _3, \\
& u_3=-\lambda _1\lambda _2\lambda _3, \\
& \mu _i=v_1\lambda _i^2+v_2\lambda
_i+v_3,\,\,\,\,\,\,\,\,\,\,\,\,\,\,\,\,\,\,\,\,i=1,2,3.
\end{split}
\label{126}
\end{equation}
Hence, eliminating the HF coordinates $(u,v)$ from the system (\ref{124})-(%
\ref{126}) we arrive at the explicit relation between DN coordinates $%
(\lambda ,\mu )$ and natural coordinates $(q,p)$. Unfortunately, the
formulas are too long to quote them in the text. In DN coordinates the two
Poisson structures take the form (\ref{119}) and the related separated curve 
\begin{equation}
\mu ^3-\lambda ^4=\mu (c_2\lambda +h_1)+c_1\lambda ^2+h_2\lambda +h_3.
\label{127}
\end{equation}
The implicit solutions of the system can be obtained by solving the three
decoupled first order ODEs 
\begin{equation}
\left( \frac{\partial W_i}{\partial \lambda _i}\right) ^3=\frac{\partial W_i%
}{\partial \lambda _i}(c_2\lambda _i+a_1)+(\lambda _i^4+c_1\lambda
_i^2+a_2\lambda _i+a_3),\,\,\,i=1,2,3.  \label{128}
\end{equation}
\end{example}

\section{Concluding remarks}

In this review paper we presented a multi-Hamiltonian separability theory of
finite dimensional dynamical systems, in the frame of the set of canonical
coordinates. It reveals the important fact that the multi-Hamiltonian
property of considered system is closely related to its separability.
Actually, we presented the constructive method of finding a separation
coordinates once having a bi(multi)-Hamiltonian representation of the
underlying dynamical system, written down in some natural canonical
coordinates. There is still an open question: whether each bi-Hamiltonian
chain with sufficient number of constants of motion guarantees a
separability of underlying Hamiltonian systems? Or, in other words: whether
the arbitrary degenerate Poisson pencil, written in noncanonical
coordinates, can be restricted to the nondegenerate one on a symplectic leaf
of one of the Poisson tensors from the pencil? The second important problem
is how to perform effectively such a Marden-Ratiu projection if it is
possible? Some progress in this direction was made recently \cite{fal} but
this part of separability theory still requires further investigations.

\label{lastpage}

\end{document}